\documentclass[twocolumn,showpacs,preprintnumbers,amsmath,amssymb,nofootinbib]{revtex4-1}

\usepackage[dvipdfmx]{graphicx}
\usepackage{amsfonts}
\usepackage[mathscr]{eucal}
\usepackage{dcolumn}
\usepackage{bm}
\usepackage[colorlinks=true,linkcolor=blue,citecolor=blue]{hyperref}
\usepackage{hyperref}
\usepackage{color}

\begin{document}


\newcommand{\vev}[1]{ \left\langle {#1} \right\rangle }
\newcommand{\bra}[1]{ \langle {#1} | }
\newcommand{\ket}[1]{ | {#1} \rangle }
\newcommand{\eV}{ \ {\rm eV} }
\newcommand{\KeV}{ \ {\rm keV} }
\newcommand{\MeV}{\  {\rm MeV} }
\newcommand{\GeV}{\  {\rm GeV} }
\newcommand{\TeV}{\  {\rm TeV} }
\newcommand{\1}{\mbox{1}\hspace{-0.25em}\mbox{l}}
\newcommand{\Red}[1]{{\color{red} {#1}}}

\newcommand{\lmk}{\left(}
\newcommand{\rmk}{\right)}
\newcommand{\lkk}{\left[}
\newcommand{\rkk}{\right]}
\newcommand{\lhk}{\left \{ }
\newcommand{\rhk}{\right \} }
\newcommand{\del}{\partial}
\newcommand{\la}{\left\langle}
\newcommand{\ra}{\right\rangle}
\newcommand{\half}{\frac{1}{2}}

\newcommand{\bea}{\begin{array}}
\newcommand{\eea}{\end{array}}
\newcommand{\beq}{\begin{eqnarray}}
\newcommand{\eeq}{\end{eqnarray}}

\newcommand{\dd}{\mathrm{d}}
\newcommand{\Mpl}{M_{\rm Pl}}
\newcommand{\mg}{m_{3/2}}
\newcommand{\abs}[1]{\left\vert {#1} \right\vert}
\newcommand{\mphi}{m_{\phi}}
\newcommand{\Hz}{\ {\rm Hz}}
\newcommand{\for}{\quad \text{for }}
\newcommand{\Min}{\text{Min}}
\newcommand{\Max}{\text{Max}}
\newcommand{\Kahler}{K\"{a}hler }
\newcommand{\cphi}{\varphi}
\newcommand{\Tr}{\text{Tr}}
\newcommand{\diag}{{\rm diag}}

\newcommand{\SUf}{SU(3)_{\rm f}}
\newcommand{\Upq}{U(1)_{\rm PQ}}
\newcommand{\Zpq}{Z^{\rm PQ}_3}
\newcommand{\Cpq}{C_{\rm PQ}}
\newcommand{\ubar}{u^c}
\newcommand{\dbar}{d^c}
\newcommand{\ebar}{e^c}
\newcommand{\nubar}{\nu^c}
\newcommand{\Ndw}{N_{\rm DW}}
\newcommand{\Fpq}{F_{\rm PQ}}
\newcommand{\fpq}{v_{\rm PQ}}
\newcommand{\Br}{{\rm Br}}
\newcommand{\Lag}{\mathcal{L}}
\newcommand{\Lqcd}{\Lambda_{\rm QCD}}
\newcommand{\const}{\text{const}}

\newcommand{\ji}{j_{\rm inf}}
\newcommand{\jb}{j_{B-L}}
\newcommand{\M}{M}
\newcommand{\im}{{\rm Im} }
\newcommand{\re}{{\rm Re} }
\newcommand{\cm}{\ {\rm cm} }

\def\lrf#1#2{ \left(\frac{#1}{#2}\right)}
\def\lrfp#1#2#3{ \left(\frac{#1}{#2} \right)^{#3}}
\def\lrp#1#2{\left( #1 \right)^{#2}}
\def\REF#1{Ref.~\cite{#1}}
\def\SEC#1{Sec.~\ref{#1}}
\def\FIG#1{Fig.~\ref{#1}}
\def\EQ#1{Eq.~(\ref{#1})}
\def\EQS#1{Eqs.~(\ref{#1})}
\def\blue#1{\textcolor{blue}{#1}}
\def\red#1{\textcolor{blue}{#1}}

\def\eq#1{Eq.~(\ref{#1})}

\def\order#1{\mathcal{O}(#1)}


\title{
A natural and simple UV completion of the QCD axion model
}

\author{
Masaki Yamada$^{1, 2}$
}

\author{
Tsutomu T. Yanagida$^{3, 4}$
}

\affiliation{
$^{1}$ Frontier Research Institute for Interdisciplinary Sciences, Tohoku University, Sendai, Miyagi 980-8578, Japan
}
\affiliation{
$^{2}$ Department of Physics, Tohoku University, Sendai, Miyagi 980-8578, Japan
}

\affiliation{
$^{3}$ T. D.  Lee Institute and School of Physics and Astronomy, Shanghai Jiao Tong University, 800 Dongchuan Rd, Shanghai 200240, China
}
\affiliation{$^{4}$ Kavli IPMU (WPI), UTIAS,
The University of Tokyo, 5-1-5 Kashiwanoha, Kashiwa, Chiba 277-8583, Japan
}

\date{\today}

\begin{abstract}
The novel PQ mechanism replaces the strong CP problem with some challenges in a model building. In particular, the challenges arise regarding i) the origin of an anomalous global symmetry called a PQ symmetry, ii) the scale of the PQ symmetry breaking, and iii) the quality of the PQ symmetry. In this letter, we provide a natural and simple UV completed model that addresses these challenges. Extra quarks and anti-quarks are separated by two branes in the Randall-Sundrum ${\bm R}^4 \times S^1 / {\bm Z}_2$ spacetime while a hidden SU($N_H$) gauge field condensates in the bulk. The brane separation is the origin of the PQ symmetry and its breaking scale is given by the dynamical scale of the SU($N_H$) gauge interaction. The (generalized) Casimir force of SU($N_H$) condensation stabilizes the 5th dimension, which guarantees the quality of the PQ symmetry.
\end{abstract}

\maketitle

{\bf Introduction.--}
Small parameters in the Standard Model (SM) are outstanding mysteries of particle physics
because they imply severe fine-tunings among bare parameters and/or quantum corrections.
Although the very small CP phase in the QCD sector is technically natural,
source of hadronic CP-violation typically produces ${\cal O}(10^{-4})$ threshold corrections
to the CP phase in the QCD sector in the SM.
This fine-tuning problem implies physics beyond the SM as a UV theory, such as
the QCD axion model that addresses the smallness of the CP phase ($\lesssim 10^{-10}$) by the PQ mechanism~\cite{Peccei:1977hh, Peccei:1977ur,Weinberg:1977ma,Wilczek:1977pj}.
However,
one should be careful when constructing a UV theory so that it indeed solves the problem
without any costs for additional fine-tunings of other parameters.
For example,
the QCD axion models introduce a very precise global symmetry called a PQ symmetry
that is anomalous under the SU(3)$_c$ gauge symmetry.
Such a global symmetry is expected to be broken by quantum gravity effects~\cite{Georgi:1981pu, Dine:1986bg, Giddings:1988cx, Coleman:1988tj, Gilbert:1989nq, Barr:1992qq, Kamionkowski:1992mf, Holman:1992us, Ghigna:1992iv, Dobrescu:1996jp, Banks:2010zn}.
In addition, the energy density of the axion may exceed the observed dark matter (DM) density
unless the scale of the PQ symmetry breaking is of the order $10^{11\, \text{-}\,12} \GeV$ or smaller~\cite{Preskill:1982cy, Abbott:1982af, Dine:1982ah}.
There is also a lower bound of the order $10^8 \GeV$
from the energy loss in the supernova SN 1987A~\cite{Mayle:1987as,Raffelt:1987yt}.
These introduce a new energy scale that is much smaller than the Planck scale
and much larger than the electroweak scale.
Thus, the PQ mechanism replaces the strong CP problem with the following challenges in a model building:
\begin{itemize}
\item origin of the PQ symmetry,
\item quality of the PQ symmetry,
\item scale of the PQ symmetry breaking.
\end{itemize}
Several studies proposed explanations of the quality problem.
The PQ symmetry is realized
as an accidental symmetry from discrete gauge symmetries~\cite{Chun:1992bn, BasteroGil:1997vn,Babu:2002ic,Dias:2002hz,Harigaya:2013vja,Harigaya:2015soa},
abelian gauge symmetries~\cite{Fukuda:2017ylt, Duerr:2017amf, Fukuda:2018oco, Bonnefoy:2018ibr, Ibe:2018hir},
and non-abelian gauge symmetries~\cite{Randall:1992ut,Dobrescu:1996jp,Redi:2016esr,DiLuzio:2017tjx,Lillard:2018fdt,Lee:2018yak,Gavela:2018paw,Buttazzo:2019mvl,Ardu:2020qmo, Yin:2020dfn}.
Models with an extra dimension are also proposed in this context~\cite{Cheng:2001ys,Izawa:2002qk,Fukunaga:2003sz,Choi:2003wr, Izawa:2004bi,Flacke:2006ad,Kawasaki:2015lea,Yamada:2015waa,Cox:2019rro}.
An intermediate scale can be introduced without a fine-tuning
by a dynamical symmetry breaking of a gauge symmetry that simultaneously breaks the PQ symmetry~\cite{Kim:1979if}.

In this letter, we provide a simple UV model that naturally realizes the PQ mechanism,
combining the ideas proposed in Refs.~\cite{Izawa:2002qk,Fujikura:2019oyi}.
We consider a warped ${\bm R}^4 \times S^1 / {\bm Z}_2$ spacetime,
where two branes, called IR and UV branes, are placed at the orbifold fixed points.
We separately put extra quarks $Q$ and $\bar{Q}$ into the different branes
and introduce a SU($N_H$) gauge field in the bulk.
Then the chiral symmetry, or the PQ symmetry, is guaranteed by the separation in the five-dimensional space~\cite{Izawa:2002qk,Fukunaga:2003sz,Izawa:2004bi,Kawasaki:2015lea,Yamada:2015waa}
and is spontaneously broken by the SU($N_H$) condensation.
Since the condensation scale of SU($N_H$) gauge theory is determined by dimensional transmutation,
its energy scale can be naturally as small as $10^{8\, \text{-}\,12} \GeV$~\cite{Kim:1979if}.
The size of the extra dimension (radion) is stabilized without introducing additional ingredients in the model.
The SU($N_H$) condensation energy depends on the size of the 5th dimension,
which provides a potential for the radion to be stabilized around the PQ scale~\cite{vonHarling:2017yew,Fujikura:2019oyi}.
In addition,
the warp factor ameliorates (though not completely addresses)
the electroweak hierarchy problem~\cite{Randall:1999ee}
and the cutoff scale is reduced to the PQ symmetry breaking scale rather than the Planck scale.
In summary, the answers to the above-mentioned issues in the PQ mechanisms are as follows:
\begin{itemize}
\item brane separation of $Q$ and $\bar{Q}$,
\item radion stabilization at the PQ scale,
\item dynamical scale of SU($N_H$).
\end{itemize}
Compared with the other related works with the 5th dimension~\cite{Cheng:2001ys,Izawa:2002qk,Fukunaga:2003sz,Choi:2003wr, Izawa:2004bi,Flacke:2006ad,Cox:2019rro},
our model automatically stabilizes the size of the 5th dimension.

\begin{figure}[t]
\centering
\includegraphics[width=.40\textwidth, bb=0 0 499 531]{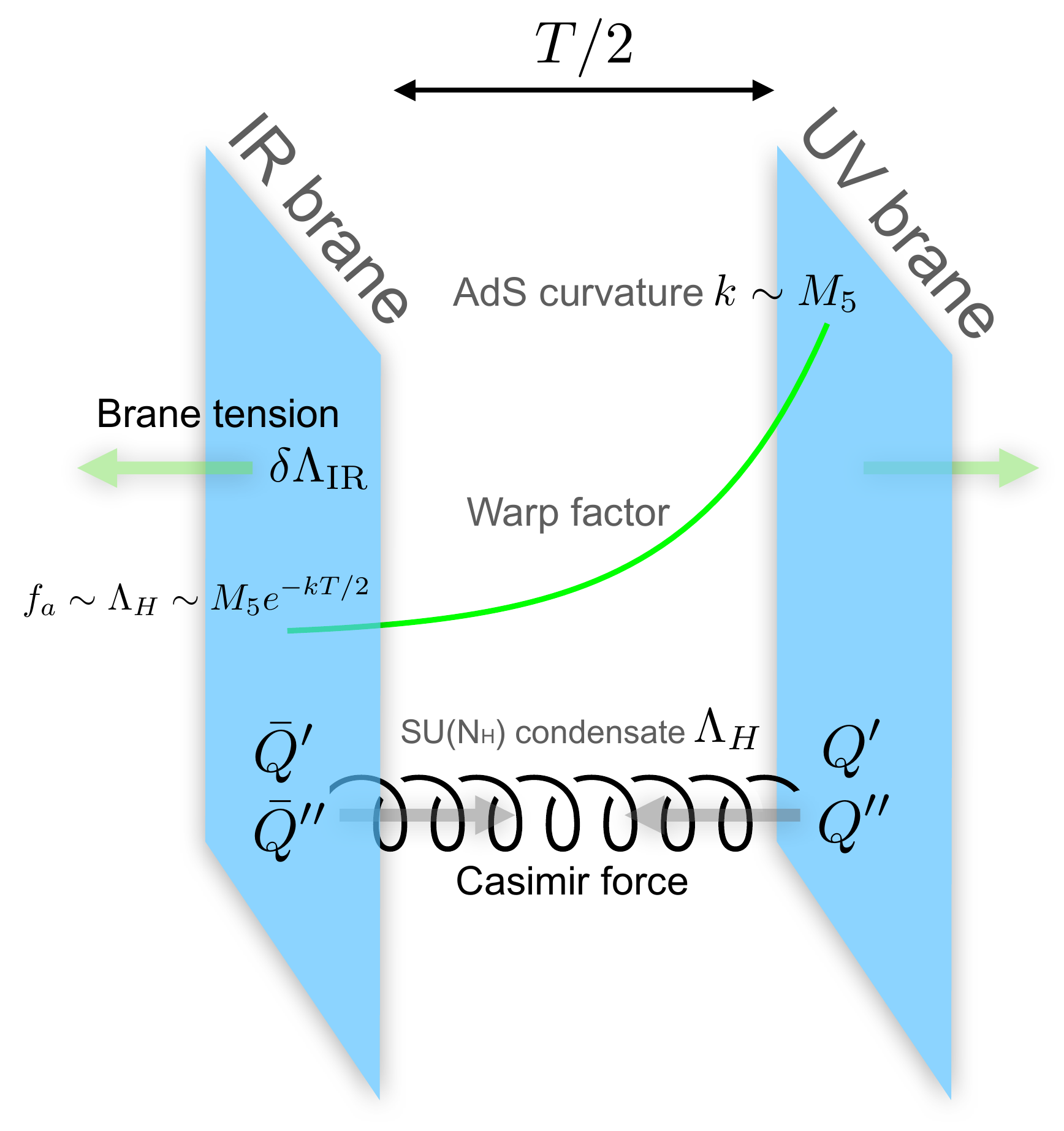}
\caption{
Schematic illustration of the model in the warped ${\bm R}^4 \times S^1 / {\bm Z}_2$ spacetime.
The horizontal line represents the 5th dimensional space.
}
  \label{fig1}
\end{figure}

\vspace{0.3cm}
{\bf QCD axion model in ${\bm R}^4 \times S^1 / {\bm Z}_2$ spacetime.--}
The model proposed in this study
is similar to the one proposed in Ref.~\cite{Izawa:2002qk} but
features a warped extra dimension~\cite{Fukunaga:2003sz}.
More specifically, we consider an SU($N_H$) gauge theory
in a warped ${\bm R}^4 \times S^1 / {\bm Z}_2$ spacetime~\cite{Randall:1999ee}.
We introduce a pair of chiral fermions $Q'({\bm 3}, {\bm N}_H)$ and $\bar{Q}'(\bar{\bm 3}, \bar{\bm N}_H)$
and $N_F-3$ pairs of chiral fermions $Q_i''({\bm 1}, {\bm N}_H)$ and $\bar{Q}_i''({\bm 1}, \bar{\bm N}_H)$,
where the arguments represent how the fermions transform under the SU(3)$_c \times$SU($N_H$) gauge group.
We collectively denote the fermions as $Q$ ($\supset Q', Q''$) and anti-fermions as $\bar{Q}$ ($\supset \bar{Q}', \bar{Q}''$). Namely, if we explicitly write the flavor index,
they are given by
\beq
 &&Q_i = Q'_a \delta_{a i} \qquad (i = 1, 2, 3),
 \\
 &&Q_i = Q''_{i -3} \qquad (i \ge 4),
\eeq
and similarly for $\bar{Q}$,
where $a$ represents the color index.
The fields $Q$ and $\bar{Q}$ are localized on UV and IR branes, respectively,
while the SU($N_H$) gauge field lives in the bulk (see Fig.~\ref{fig1}).
The standard model (SM) particles are localized on the IR brane.
The metric is given by
\begin{align}
ds^2
= e^{-2kT(x)\left.| y \right.|} g_{\mu\nu} dx^\mu dx^\nu - T^2 (x) dy^2,
\label{eq:RS metric}
\end{align}
where
$\mu, \nu$ run from $0$ to $3$,
$g_{\mu \nu}$ is the 4D induced metric, $y$ $\in (-1/2,  1/2)$ represents the coordinate for the $5$th dimension
with ${\bold Z}_2$ symmetry $y \leftrightarrow -y$,
and $k$ is the AdS curvature.
The parameter $T(x)$ represents the size of the extra dimension.
We denote $T_0$ as the size at present and
define the radion field by $\mu \equiv k e^{-k T(x)/2}$.

The Lagrangian in the bulk is given by
\beq
{\cal L}_{\rm bulk} &&= \frac{1}{2} M_5^3 R - V_5 -\frac{1}{4g_{c5}^2} G_{AB} G^{AB}
\nonumber\\
&&~~~~ -\frac{1}{4g_{h5}^2} F_{AB} F^{AB} + {\cal L}_{\rm CS},
\label{eq:E-H action}
\eeq
where
$M_5$ and $R$ are the 5D Planck mass and the Ricci scalar,
$V_5$ ($=-6 M_5^3 k^2$) is a bulk cosmological constant,
$G_{AB}$ and $F_{AB}$ are the 5D gauge field strengths of SU(3)$_c$ and SU($N_H$),
$g_{c5}$ and $g_{h5}$ are their gauge coupling constants,
and ${\cal L}_{\rm CS}$ is a Chern-Simons term that cancels gauge anomalies~\cite{Izawa:2002qk}.%
\footnote{If one considers a grand unified theory (GUT), all SM gauge fields, including U(1)$_Y$, must live in the bulk. This does not affect our discussion.}

The Lagrangians on the IR and UV branes are given by
\beq
&& {\cal L}_{\rm IR} = {\cal L}_{\bar Q}
- \frac{\tau_{c,{\rm IR}}}{4} G_{\mu \nu} G^{\mu \nu}
- \frac{\tau_{h,{\rm IR}}}{4} F_{\mu \nu} F^{\mu \nu}
- V_{\rm IR}
 \\
&& {\cal L}_{\rm UV} = {\cal L}_Q
- \frac{\tau_{c,{\rm UV}}}{4} G_{\mu \nu} G^{\mu \nu}
- \frac{\tau_{h,{\rm UV}}}{4} F_{\mu \nu} F^{\mu \nu}
 - V_{\rm UV},
\eeq
respectively,
where ${\cal L}_Q$ and ${\cal L}_{\bar Q}$ are kinetic terms of $Q$ and $\bar{Q}$
and their higher-dimensional terms that will be discussed later.
We include localized kinetic terms for the gauge fields into both branes.
We omit the SM Lagrangian that may be localized on the IR brane for notational simplicity.
The IR and UV brane tensions are rewritten as
\beq
 V_{\rm IR} = - 6 M_5^3 k + \delta V_{\rm IR},
 \quad
 V_{\rm UV} = 6 M_5^3 k + \delta  V_{\rm UV},
 \label{Lambda_IR}
\eeq
respectively.

Every mass parameter on the IR brane is exponentially suppressed by the warp factor $e^{-kT_0 /2}$
when measured with the 4D Einstein metric.
The hierarchy problem is then ameliorated for $k T_0 \gg 1$~\cite{Randall:1999ee}.
However, it is not our primary motivation to consider the warped extra-dimension.
The 4D (reduced) Planck scale is given by $M_{\rm Pl}^2 = M_5^3 k^{-1} (1- e^{-k T_0})$.

\vspace{0.3cm}
{\bf Radion stabilization.--}
Next, we explain the radion stabilization in our model, following Refs.~\cite{Fujikura:2019oyi} and \cite{vonHarling:2017yew}.
We can consider a four-dimensional effective field theory by
the KK decomposition and integrating out heavier particle than $\mu$.
We then obtain a 4D effective action for the zero-mode gauge field of SU($N_H$) with a gauge coupling of~\cite{Agashe:2002bx,Csaki:2007ns}
\begin{align}
\frac{1}{g_{h4}^2 } =
-\frac{b_g}{8\pi^2} \log\left(\frac{k}{\mu}\right)
-\frac{b_{\rm UV}}{8\pi^2} \log\left(\frac{k}{E}\right)
\nonumber\\
-\frac{b_{\rm IR}}{8\pi^2} \log\left(\frac{\mu}{E}\right)
+\tau_{h,{\rm UV}} + \tau_{h,{\rm IR}}
\end{align}
at the energy scale of $E$ ($\lesssim \mu$).
Here $b_g \equiv - 8 \pi^2 / (k g_{h5}^2)$,
$b_{\rm UV}= 11 N_H /3 - N_F/3$, and
$b_{\rm IR} = - N_F/3$.
When $b_{\rm UV} + b_{\rm IR} >0$, the SU($N_H$) gauge interaction is asymptotically free
and is confined at the energy scale of
\beq
\Lambda_{H} (\mu)
&=& \left(  k^{b_{\rm UV}} \mu^{b_{\rm IR}} e^{-8 \pi^2 (\tau_{\rm IR} + \tau_{\rm UV}) } \left(\frac{\mu}{k}\right)^{-b_g}
\right)^{1/(b_{\rm UV}+b_{\rm IR})}
\nonumber\\
&\equiv& \Lambda_{H,0} \left( \frac{\mu}{\mu_0}\right)^n,
\label{eq:radion dependent confinement scale}
\eeq
for $\Lambda_{H} (\mu) \lesssim \mu$,
where $\mu_0 \equiv k e^{- kT_0 /2}$ and
$n\equiv (b_{\rm IR} - b_g)/(b_{\rm UV} + b_{\rm IR})$.
Here we explicitly express the $\mu$ dependence from the dynamical scale.

The vacuum energy of the condensation
can be obtained from the dimensional analysis and is given by
\beq
 \frac{1}{4} \la T^\mu_\mu \ra
 &=& -\frac{1}{4} \frac{b_{\rm UV} + b_{\rm IR}}{32\pi^2} \la F^{(0)}_{\mu\nu} F^{(0) \mu\nu} \ra
 \\
 &\sim& -\frac{1}{4} \frac{b_{\rm UV} + b_{\rm IR}}{32\pi^2} (4\pi)^2 \Lambda^4_{H} (\mu).
\eeq
Note that this estimation is supported by the lattice simulation for the SM QCD~\cite{vonHarling:2017yew}.
The result depends on $\mu$ via the Beta function.
The effective action of the radion field $\mu$ is then read as
\begin{align}
&S_{\rm radion} = \int d^4 x \left[ \, 3 \lmk \frac{M_5}{k} \rmk^3  \left( \partial \mu(x)\right)^2 - V(\mu) \right], \label{radionkinetic} \\[1.5ex]
&V(\mu)= \delta V_{\rm UV} + \frac{ \delta V_{\rm IR}}{k^4} \mu^4 -\frac{b_{\rm UV} + b_{\rm IR}}{8} \Lambda^4_{H, 0} \left(\frac{\mu}{\mu_0}\right)^{4n}. \label{eq:radion effective potential}
\end{align}
We consider a large $M_5 / k$ so that quantum gravity effects are negligible.
According to the naive dimensional analysis~\cite{Agashe:2007zd}, it requires~\cite{vonHarling:2017yew}
\beq
\label{Nmin}
 2\pi (M_5 / k)^{3/2} \gtrsim 4 \cdot 5^{3/4} / \sqrt{3 \pi}  ~~\leftrightarrow ~~M_5 \gtrsim 0.6\, k.
\eeq
The cosmological constant at the minimum can be made vanishingly small by choosing $\delta V_{\rm UV}$ appropriately, which is the only fine-tuning we require in our model.

The branes repel with each other due to the IR brane tension.
At the same time,
they are attracted by the condensation energy of SU($N_H$).
The radion VEV is determined by the balance between these forces.
The VEV and the mass of the radion are given by~\cite{Fujikura:2019oyi}
\begin{align}
 &\mu_0 = \lmk \frac{n (b_{\rm UV} + b_{\rm IR} )k^4}{8 \delta V_{\rm IR}} \rmk^{1/4} \Lambda_{H,0},
 \label{mu_min}
 \\
 &m_{\rm radion}^2 = \lmk 1-n \rmk  \left( \frac{32\pi^2}{3N^2}\right)
 \lmk \frac{\delta V_{\rm IR}}{k^4} \rmk
  \mu_0^2 \, , \label{eq:radion mass}
\end{align}
for $n < 1$,
where we implicitly assume that $\mu_0 \gtrsim \Lambda_{H,0}$ since otherwise \eq{eq:radion dependent confinement scale} cannot be used.
From \eq{Lambda_IR}, we expect $\delta V_{\rm UV} \sim M_5^3 k$ or a little bit smaller
to avoid a fine-tuning in the IR brane tension.
We also expect that $M_5 / k$ satisfies \eq{Nmin} but is not significantly larger than unity.
Then the expression in the parentheses in \eq{mu_min} is of order (but is larger than) unity.
Hence we obtain $e^{- k T_0/2} = \mu_0 / k \sim \Lambda_{H,0} / k$.
As we will see shortly, we consider $\Lambda_{H,0} /N_H \sim f_a$ ($= {\cal O}(10^{8\, \text{-} \,12}) \GeV$),
so that the warp factor is estimated as $e^{- k T_0/2} \sim 10^{-(6\, \text{-} \,10)}$ (i.e., $k T_0 \simeq 28\, \text{-} \, 46$).
This is not small enough to completely address the hierarchy problem between the electroweak
and the Planck scales but ameliorates it by a factor of the order $N_H f_a / k$.
This is another advantage of our model. It does not only address the issues in the QCD axion model
but also ameliorates the hierarchy problem.

\vspace{0.3cm}
{\bf Origin of PQ symmetry.--}
Now we shall explain how the PQ mechanism is realized at an intermediate scale and how the quality of the symmetry is
guaranteed in the model.
In this explanation, we follow Ref.~\cite{Izawa:2002qk}.
For a moment, let us consider SU($N_H$) gauge interaction
and omit the SU(3)$_c$ gauge interaction.
Then there are
$N_F$ pairs of chiral fermions, $Q_i$ and $\bar{Q}_i$, in the SU($N_H$) gauge theory.
Since $Q_i$ and $\bar{Q}_i$ are
separately placed on different branes,
the operators involving both $Q_i$ and $\bar{Q}_i$ are exponentially suppressed
and the model possesses an approximate $U(N_F)_V \times U(N_F)_A$ flavor symmetry,
where $U(N_F)_V$ and $U(N_F)_A$ represent the vector and axial transformations, respectively.
In particular, the vector mass term $M_{Q_i} Q_i \bar{Q}_i$ is suppressed as $M_{Q_i} \propto e^{- c M_5 T_0 }$ with $c = {\cal O}(1)$.
However, because of the chiral anomaly of the SU($N_H$) gauge theory, $U(N_F)_A$ is broken to $SU(N_F)_A$.
In addition, one may write the following higher-dimensional operators on each brane that explicitly break the flavor symmetry:
\beq
 && {\cal L}_Q \supset \frac{y_Q}{M_5^{3N_H - 4}} \lmk Q \rmk^{2N_H}  + {\rm H. c.},
 \nonumber\\
 && {\cal L}_{\bar{Q}} \supset \frac{y_{\bar{Q}}}{M_5^{3N_H - 4}} \lmk \bar{Q} \rmk^{2N_H}  + {\rm H. c.},
 \label{PQ breaking terms1}
\eeq
for an odd $N_H$,
where $y_Q$ and $y_{\bar{Q}}$ are coupling constants
and we omit the flavor indices for notational simplicity.
For an even $N_H$, we obtain similar terms with a replacement of $N_H \to N_H/2$.
One can forbid these terms by making $Q$ and $\bar{Q}$ charged under U(1)$_Y$ and/or U(1)$_{B-L}$.
Even if they are not forbidden by gauge symmetries,
one can check that they do not spoil the PQ mechanism for a large $N_H$ and/or a small $f_a$ (see Ref.~\cite{Fukunaga:2003sz} for detail).

As we discussed, the SU($N_H$) gauge interaction confines at the energy scale of $\Lambda_{H,0}$.
Then the chiral condensate develops such as
\beq
 \la Q_i \tilde{\bar{Q}}_j \ra \sim
 \Lambda_{H,0}^3 \delta_{ij}
 \label{condensation}
\eeq
where $\tilde{\bar{Q}}_i$ $(\equiv e^{-3k T_0/4} \bar{Q}_i)$ is a rescaled field of $\bar{Q}_i$ to canonicalize the kinetic term in the four-dimensional effective field theory.
As a result, the $U(N_F)_V \times SU(N_F)_A$ flavor symmetry is spontaneously broken to the $U(N_F)_V$ symmetry.
The number of composite NG bosons would then be $N_F^2-1$, where a factor of $-1$ comes from a massive pseudo-NG boson due to the chiral anomaly of the SU($N_H$) gauge theory.

Now let us add the SU(3)$_c$ gauge interaction,
where the SU(3) ($\in$ SU($N_F$)$_V$) flavor symmetry is promoted to the SU(3)$_c$ gauge symmetry.
The $U(N_F)_V \times SU(N_F)_A$ flavor symmetry is then explicitly broken by $SU(3)_c$ gauge interactions
down to $U(1)_{\rm PQ} \times U(N_F-3)_V \times SU(N_F-3)_A$,
where $U(N_F-3)_V$ and $SU(N_F-3)_A$ are the vector and axial transformation on $Q''({\bm 1},{\bm N_H})$ and $\bar{Q}''({\bm 1},{\bm N_H})$,
respectively.
The U(1)$_{\rm PQ}$ symmetry is defined by
\beq
\left\{
\bea{ll}
 Q' (\bar{Q}') \to e^{i \alpha/3} Q'  (\bar{Q}'),
\\
 Q''  (\bar{Q}'') \to e^{-i \alpha/(N_F - 3)} Q'' (\bar{Q}'').
\eea
\right.
\eeq
This is anomalous under the SU(3)$_c$ gauge interaction.
Accordingly,
the associated pseudo-NG boson is identified as the axion.
In summary,
$U(1)_{\rm PQ} \times U(N_F-3)_V \times SU(N_F-3)_A$
is spontaneously broken to $U(N_F-3)_V$ by the condensation of SU($N_H$) (see \eq{condensation}).
Hence,
there are $(N_F - 3)^2 -1$ NG bosons as well as the axion in the effective theory below the condensation scale.
The PQ symmetry breaking scale is therefore identified as the condensation scale $\Lambda_{H,0}$.
Since the condensation scale is determined by the dynamical scale of SU($N_H$),
the smallness of its energy scale (compared to the fundamental scales, e.g., the Planck scale)
is explained by the dimensional transmutation~\cite{Kim:1979if}.

Now we go back to the symmetry-breaking operator,
which particularly breaks the PQ symmetry.
The vector mass $M_{Q_i} Q_i \bar{Q}_i$ induces a shift in the strong CP phase of the order $M_{Q_i} e^{-kT_0/2} f_a/m_a^2$, where $m_a$ is the axion mass at the low energy and $f_a$ ($\simeq \Lambda_{H,0}/ N_H$) is the axion decay constant.
Requiring $\Delta a / f_a \lesssim 10^{-10}$ to solve the strong CP problem,
we find that $c M_5 T_0 \gtrsim 144 + 4 \ln ( f_a / 10^{12} \GeV)$ is sufficient.
Since $M_5 \gtrsim k$ and $k T_0 \sim 28\, \text{-}\,46$,
the strong CP phase is sufficiently small for $c \gtrsim 3\, \text{-}\,5$.
Thus the quality of the PQ symmetry is explained by the brane separation in our model~\cite{Izawa:2002qk}.

\vspace{0.3cm}
{\bf Cosmological scenario.--}
Finally, we explain the cosmological scenario of our model.
Since $Q'({\bm 3}, {\bm N_H})$ and $\bar{Q}'(\bar{\bm 3}, \bar{\bm N_H})$ contribute
to the $U(1)_{\rm PQ} \times SU(3)_c \times SU(3)_c$ chiral anomaly,
the domain wall number of the axion is equal to $N_H$ and hence $f_a \sim \Lambda_{H,0} / N_H$.
To avoid the inhomogeneous Universe due to the production of stable domain walls,
we consider the pre-inflationary PQ symmetry breaking scenario.
Then
the energy density of the coherent oscillation of the axion is determined by the misalignment mechanism
such as~\cite{Wantz:2009it, Kawasaki:2014sqa}
\beq
 \Omega_a h^2
 \simeq
 0.12 \, \theta_{\rm ini}^2
 \lmk \frac{f_a}{10^{12} \GeV} \rmk^{1.165},
  \label{axion abundance}
\eeq
where $h$ is the Hubble parameter in units of $100 \ {\rm km/s/Mpc}$
and $\theta_{\rm ini}$ is the initial misalignment angle.
The axion decay constant $f_a$ should be of order $10^{12} \GeV$ to explain all DM unless $\theta_{\rm ini}$ is small.
Such a ``small" energy scale of $f_a$ is naturally realized in our model due to the dimensional transmutation.

The axion acquires quantum fluctuations during inflation, whose amplitude is proportional to the energy scale of inflation.
Those modes result in isocurvature density perturbations~\cite{Axenides:1983hj, Seckel:1985tj, Turner:1990uz}.
The constraint by the Planck collaboration implies that the energy scale of inflation has to be smaller than of order $10^7 \GeV$~\cite{Ade:2015lrj}
if the axion is all DM.
In fact, a small energy scale of inflation may be a natural consequence of the anthropic landscape~\cite{Kawasaki:2015mvm},
where inflations occur at infinitely many vacuum states with different energy scales.
It is a possible move to a different vacuum state
by a quantum tunneling process.
Since the rate of upward tunneling (namely Hawking-Moss transition) is strongly suppressed compared with that of downward tunneling, it would take a journey to a lower energy scale.
Eventually, there is an option to go through a slow-roll region and reach a habitable vacuum with a vanishingly small vacuum energy.
Accordingly, we expect that the last slow-roll inflation is a small-scale one,
which is consistent with the isocurvature constraint.

Since we consider the pre-inflationary PQ symmetry breaking scenario,
the reheating temperature (and the maximal temperature) should be lower than the condensation scale of SU($N_H$), namely $N_H f_a$.
The colored would-be NG bosons acquire effective masses
of the dynamical scale order due to the radiative correction from the SU(3)$_c$ gauge interaction
and are not produced after inflation.
If one considered a scenario where they are produced,
$\bar{Q}'$ would have be made charged under U(1)$_Y$ so that the triplet NG bosons have the same SM charges with the SM down quarks and
can decay into SM particles.
If the reheating temperature after inflation is as high as of the order $f_a / 10$,
the $(N_F - 3)^2 - 1$ massless singlet NG bosons as well as the axion are thermalized~\cite{Masso:2002np, Kawasaki:2015ofa}
and contribute to the energy density of the Universe as dark radiation~\cite{Turner:1986tb, Nakayama:2010vs, Weinberg:2013kea}.
The resulting abundance is conveniently expressed by the effective neutrino number, which is given by~\cite{Salvio:2013iaa, Kawasaki:2015ofa}
\beq
 N_{\rm eff} \simeq N_{\rm eff}^{({\rm SM})} + 0.027 \times (N_F - 3)^2,
\eeq
where $N_{\rm eff}^{({\rm SM})}$ ($\simeq 3.046$) is the SM prediction.
Even if $N_F = 4$, the deviation from the SM prediction would be measured
by CMB-S4 in the future~\cite{Wu:2014hta} (see also Ref.~\cite{Abazajian:2013oma}).

\vspace{0.3cm}
{\bf Summary and discussion.--}
Summarizing the conditions on our model parameters,
we require
$M_5^3 / k = \Mpl^2$ ($\simeq 2.4 \times 10^{18} \GeV$),
$M_5/ k \gtrsim 0.6$,
$T_0^{-1} = k / (28 \, \text{-} \, 46$),
and $144 / (M_5 T_0) = {\cal O}(1)$. These are satisfied, e.g.,
$k = 2 \times 10^{18} \GeV$, $M_5 = 2.3 \times 10^{18} \GeV$, $T_0^{-1} = 4 \, \text{-} \, 7 \times 10^{16} \GeV$.
There are no small parameters nor fine tunings in our model, except for the notorious fine tuning on the vanishing 4-dimensional cosmological constant.
The intermediate PQ scale, $10^{8\, \text{-} \, 12} \GeV$,
is realized by the dimensional transmutation.
It is automatically close to the IR-brane-cutoff scale, namely the radion VEV.
This implies that
if one wants to consider a GUT, the PQ scale should be as large as the GUT scale.

One can consider the case with $f_a \lesssim 10^{11} \GeV$,
where the axion cannot explain all DM.
The constraint on the energy scale of inflation implied to avoid isocurvature density perturbations is not applicable here.
In fact, such a small $f_a$ is favored in our scenario
in light of the hierarchy problem
because the warp factor decreases with $f_a$.
Although the hierarchy problem is not completely solved,
it is ameliorated by many orders of magnitude
due to the Randall-Sundrum mechanism.
Helioscopes can search for solar axions with a relatively small $f_a$.
The sensitivity of the IAXO experiment is expected to reach
$N_H f_a \sim 10^9 \GeV$ in the future~\cite{Armengaud:2014gea}.

Our stabilization mechanism is based on the fact that
the condensation energy of SU($N_H$) gauge field depends on the size of the extra dimension via its beta function. This may be regarded as a generalized Casimir energy as it is the vacuum energy in the SU($N_H$) gauge theory and depends on the distance between the branes (boundaries). In the same spirit, the cosmological constant in the bulk as well as the brane tensions may also be regarded as a generalized Casimir energy.
It was known that Casimir forces are attractive if the boundary conditions on the two boundaries respect interchange symmetry~\cite{Casimir:1948dh,Lifshitz:1956zz,Kenneth:2006vr}. Recently, Jiang and Wilczek found a loophole of this theorem by inserting an intermediate chiral material in the bulk~\cite{Jiang:2018ivv}. Their result implies that Casimir forces can be repulsive if the medium in the bulk does not respect the interchange symmetry.
A similar logic may apply to the generalized Casimir energy, namely the cosmological constant, brane tensions, and condensation energy. The gauge field has a symmetric configuration, which results in an attractive force from the condensation energy. On the other hand, the cosmological constant in the bulk implies that the brane tensions should not be symmetric to satisfy the Einstein equation. As a result, the interchange symmetry is broken in the presence of the cosmological constant as well as the brane tensions and thus we can generate either attractive or repulsive forces.
Combining these results, the size of the extra dimension is stabilized by the balance between the attractive and repulsive Casimir forces of SU($N_H$) and the cosmological constant.

%
\vspace{0.2cm}
{\bf Acknowledgments.--}
M.\,Y.\ was supported by the Leading Initiative for Excellent Young Researchers, Ministry of Education, Culture, Sports, Science and Technology (MEXT), Japan, and by JSPS KAKENHI Grant Number JP20K22344.
T.~T.~Y. was supported in part by the China Grant for Talent Scientific Start-Up Project and the JSPS Grant-in-Aid for Scientific Research No.~16H02176, No.~17H02878, and No.~19H05810 and by World Premier International Research Center Initiative (WPI Initiative), MEXT, Japan.
%

\vspace{1cm}

\bibliography{references}

\begin{thebibliography}{76}%
\makeatletter
\providecommand \@ifxundefined [1]{%
 \@ifx{#1\undefined}
}%
\providecommand \@ifnum [1]{%
 \ifnum #1\expandafter \@firstoftwo
 \else \expandafter \@secondoftwo
 \fi
}%
\providecommand \@ifx [1]{%
 \ifx #1\expandafter \@firstoftwo
 \else \expandafter \@secondoftwo
 \fi
}%
\providecommand \natexlab [1]{#1}%
\providecommand \enquote  [1]{``#1''}%
\providecommand \bibnamefont  [1]{#1}%
\providecommand \bibfnamefont [1]{#1}%
\providecommand \citenamefont [1]{#1}%
\providecommand \href@noop [0]{\@secondoftwo}%
\providecommand \href [0]{\begingroup \@sanitize@url \@href}%
\providecommand \@href[1]{\@@startlink{#1}\@@href}%
\providecommand \@@href[1]{\endgroup#1\@@endlink}%
\providecommand \@sanitize@url [0]{\catcode `\\12\catcode `\$12\catcode
  `\&12\catcode `\#12\catcode `\^12\catcode `\_12\catcode `\%12\relax}%
\providecommand \@@startlink[1]{}%
\providecommand \@@endlink[0]{}%
\providecommand \url  [0]{\begingroup\@sanitize@url \@url }%
\providecommand \@url [1]{\endgroup\@href {#1}{\urlprefix }}%
\providecommand \urlprefix  [0]{URL }%
\providecommand \Eprint [0]{\href }%
\providecommand \doibase [0]{http://dx.doi.org/}%
\providecommand \selectlanguage [0]{\@gobble}%
\providecommand \bibinfo  [0]{\@secondoftwo}%
\providecommand \bibfield  [0]{\@secondoftwo}%
\providecommand \translation [1]{[#1]}%
\providecommand \BibitemOpen [0]{}%
\providecommand \bibitemStop [0]{}%
\providecommand \bibitemNoStop [0]{.\EOS\space}%
\providecommand \EOS [0]{\spacefactor3000\relax}%
\providecommand \BibitemShut  [1]{\csname bibitem#1\endcsname}%
\let\auto@bib@innerbib\@empty
\bibitem [{\citenamefont {Peccei}\ and\ \citenamefont
  {Quinn}(1977{\natexlab{a}})}]{Peccei:1977hh}%
  \BibitemOpen
  \bibfield  {author} {\bibinfo {author} {\bibfnamefont {R.}~\bibnamefont
  {Peccei}}\ and\ \bibinfo {author} {\bibfnamefont {H.~R.}\ \bibnamefont
  {Quinn}},\ }\href {\doibase 10.1103/PhysRevLett.38.1440} {\bibfield
  {journal} {\bibinfo  {journal} {Phys. Rev. Lett.}\ }\textbf {\bibinfo
  {volume} {38}},\ \bibinfo {pages} {1440} (\bibinfo {year}
  {1977}{\natexlab{a}})}\BibitemShut {NoStop}%
\bibitem [{\citenamefont {Peccei}\ and\ \citenamefont
  {Quinn}(1977{\natexlab{b}})}]{Peccei:1977ur}%
  \BibitemOpen
  \bibfield  {author} {\bibinfo {author} {\bibfnamefont {R.}~\bibnamefont
  {Peccei}}\ and\ \bibinfo {author} {\bibfnamefont {H.~R.}\ \bibnamefont
  {Quinn}},\ }\href {\doibase 10.1103/PhysRevD.16.1791} {\bibfield  {journal}
  {\bibinfo  {journal} {Phys. Rev. D}\ }\textbf {\bibinfo {volume} {16}},\
  \bibinfo {pages} {1791} (\bibinfo {year} {1977}{\natexlab{b}})}\BibitemShut
  {NoStop}%
\bibitem [{\citenamefont {Weinberg}(1978)}]{Weinberg:1977ma}%
  \BibitemOpen
  \bibfield  {author} {\bibinfo {author} {\bibfnamefont {S.}~\bibnamefont
  {Weinberg}},\ }\href {\doibase 10.1103/PhysRevLett.40.223} {\bibfield
  {journal} {\bibinfo  {journal} {Phys. Rev. Lett.}\ }\textbf {\bibinfo
  {volume} {40}},\ \bibinfo {pages} {223} (\bibinfo {year} {1978})}\BibitemShut
  {NoStop}%
\bibitem [{\citenamefont {Wilczek}(1978)}]{Wilczek:1977pj}%
  \BibitemOpen
  \bibfield  {author} {\bibinfo {author} {\bibfnamefont {F.}~\bibnamefont
  {Wilczek}},\ }\href {\doibase 10.1103/PhysRevLett.40.279} {\bibfield
  {journal} {\bibinfo  {journal} {Phys. Rev. Lett.}\ }\textbf {\bibinfo
  {volume} {40}},\ \bibinfo {pages} {279} (\bibinfo {year} {1978})}\BibitemShut
  {NoStop}%
\bibitem [{\citenamefont {Georgi}\ \emph {et~al.}(1981)\citenamefont {Georgi},
  \citenamefont {Hall},\ and\ \citenamefont {Wise}}]{Georgi:1981pu}%
  \BibitemOpen
  \bibfield  {author} {\bibinfo {author} {\bibfnamefont {H.~M.}\ \bibnamefont
  {Georgi}}, \bibinfo {author} {\bibfnamefont {L.~J.}\ \bibnamefont {Hall}}, \
  and\ \bibinfo {author} {\bibfnamefont {M.~B.}\ \bibnamefont {Wise}},\ }\href
  {\doibase 10.1016/0550-3213(81)90433-8} {\bibfield  {journal} {\bibinfo
  {journal} {Nucl. Phys. B}\ }\textbf {\bibinfo {volume} {192}},\ \bibinfo
  {pages} {409} (\bibinfo {year} {1981})}\BibitemShut {NoStop}%
\bibitem [{\citenamefont {Dine}\ and\ \citenamefont
  {Seiberg}(1986)}]{Dine:1986bg}%
  \BibitemOpen
  \bibfield  {author} {\bibinfo {author} {\bibfnamefont {M.}~\bibnamefont
  {Dine}}\ and\ \bibinfo {author} {\bibfnamefont {N.}~\bibnamefont {Seiberg}},\
  }\href {\doibase 10.1016/0550-3213(86)90043-X} {\bibfield  {journal}
  {\bibinfo  {journal} {Nucl. Phys. B}\ }\textbf {\bibinfo {volume} {273}},\
  \bibinfo {pages} {109} (\bibinfo {year} {1986})}\BibitemShut {NoStop}%
\bibitem [{\citenamefont {Giddings}\ and\ \citenamefont
  {Strominger}(1988)}]{Giddings:1988cx}%
  \BibitemOpen
  \bibfield  {author} {\bibinfo {author} {\bibfnamefont {S.~B.}\ \bibnamefont
  {Giddings}}\ and\ \bibinfo {author} {\bibfnamefont {A.}~\bibnamefont
  {Strominger}},\ }\href {\doibase 10.1016/0550-3213(88)90109-5} {\bibfield
  {journal} {\bibinfo  {journal} {Nucl. Phys. B}\ }\textbf {\bibinfo {volume}
  {307}},\ \bibinfo {pages} {854} (\bibinfo {year} {1988})}\BibitemShut
  {NoStop}%
\bibitem [{\citenamefont {Coleman}(1988)}]{Coleman:1988tj}%
  \BibitemOpen
  \bibfield  {author} {\bibinfo {author} {\bibfnamefont {S.~R.}\ \bibnamefont
  {Coleman}},\ }\href {\doibase 10.1016/0550-3213(88)90097-1} {\bibfield
  {journal} {\bibinfo  {journal} {Nucl. Phys. B}\ }\textbf {\bibinfo {volume}
  {310}},\ \bibinfo {pages} {643} (\bibinfo {year} {1988})}\BibitemShut
  {NoStop}%
\bibitem [{\citenamefont {Gilbert}(1989)}]{Gilbert:1989nq}%
  \BibitemOpen
  \bibfield  {author} {\bibinfo {author} {\bibfnamefont {G.}~\bibnamefont
  {Gilbert}},\ }\href {\doibase 10.1016/0550-3213(89)90097-7} {\bibfield
  {journal} {\bibinfo  {journal} {Nucl. Phys. B}\ }\textbf {\bibinfo {volume}
  {328}},\ \bibinfo {pages} {159} (\bibinfo {year} {1989})}\BibitemShut
  {NoStop}%
\bibitem [{\citenamefont {Barr}\ and\ \citenamefont
  {Seckel}(1992)}]{Barr:1992qq}%
  \BibitemOpen
  \bibfield  {author} {\bibinfo {author} {\bibfnamefont {S.~M.}\ \bibnamefont
  {Barr}}\ and\ \bibinfo {author} {\bibfnamefont {D.}~\bibnamefont {Seckel}},\
  }\href {\doibase 10.1103/PhysRevD.46.539} {\bibfield  {journal} {\bibinfo
  {journal} {Phys. Rev. D}\ }\textbf {\bibinfo {volume} {46}},\ \bibinfo
  {pages} {539} (\bibinfo {year} {1992})}\BibitemShut {NoStop}%
\bibitem [{\citenamefont {Kamionkowski}\ and\ \citenamefont
  {March-Russell}(1992)}]{Kamionkowski:1992mf}%
  \BibitemOpen
  \bibfield  {author} {\bibinfo {author} {\bibfnamefont {M.}~\bibnamefont
  {Kamionkowski}}\ and\ \bibinfo {author} {\bibfnamefont {J.}~\bibnamefont
  {March-Russell}},\ }\href {\doibase 10.1016/0370-2693(92)90492-M} {\bibfield
  {journal} {\bibinfo  {journal} {Phys. Lett. B}\ }\textbf {\bibinfo {volume}
  {282}},\ \bibinfo {pages} {137} (\bibinfo {year} {1992})},\ \Eprint
  {http://arxiv.org/abs/hep-th/9202003} {arXiv:hep-th/9202003} \BibitemShut
  {NoStop}%
\bibitem [{\citenamefont {Holman}\ \emph {et~al.}(1992)\citenamefont {Holman},
  \citenamefont {Hsu}, \citenamefont {Kephart}, \citenamefont {Kolb},
  \citenamefont {Watkins},\ and\ \citenamefont {Widrow}}]{Holman:1992us}%
  \BibitemOpen
  \bibfield  {author} {\bibinfo {author} {\bibfnamefont {R.}~\bibnamefont
  {Holman}}, \bibinfo {author} {\bibfnamefont {S.~D.}\ \bibnamefont {Hsu}},
  \bibinfo {author} {\bibfnamefont {T.~W.}\ \bibnamefont {Kephart}}, \bibinfo
  {author} {\bibfnamefont {E.~W.}\ \bibnamefont {Kolb}}, \bibinfo {author}
  {\bibfnamefont {R.}~\bibnamefont {Watkins}}, \ and\ \bibinfo {author}
  {\bibfnamefont {L.~M.}\ \bibnamefont {Widrow}},\ }\href {\doibase
  10.1016/0370-2693(92)90491-L} {\bibfield  {journal} {\bibinfo  {journal}
  {Phys. Lett. B}\ }\textbf {\bibinfo {volume} {282}},\ \bibinfo {pages} {132}
  (\bibinfo {year} {1992})},\ \Eprint {http://arxiv.org/abs/hep-ph/9203206}
  {arXiv:hep-ph/9203206} \BibitemShut {NoStop}%
\bibitem [{\citenamefont {Ghigna}\ \emph {et~al.}(1992)\citenamefont {Ghigna},
  \citenamefont {Lusignoli},\ and\ \citenamefont {Roncadelli}}]{Ghigna:1992iv}%
  \BibitemOpen
  \bibfield  {author} {\bibinfo {author} {\bibfnamefont {S.}~\bibnamefont
  {Ghigna}}, \bibinfo {author} {\bibfnamefont {M.}~\bibnamefont {Lusignoli}}, \
  and\ \bibinfo {author} {\bibfnamefont {M.}~\bibnamefont {Roncadelli}},\
  }\href {\doibase 10.1016/0370-2693(92)90019-Z} {\bibfield  {journal}
  {\bibinfo  {journal} {Phys. Lett. B}\ }\textbf {\bibinfo {volume} {283}},\
  \bibinfo {pages} {278} (\bibinfo {year} {1992})}\BibitemShut {NoStop}%
\bibitem [{\citenamefont {Dobrescu}(1997)}]{Dobrescu:1996jp}%
  \BibitemOpen
  \bibfield  {author} {\bibinfo {author} {\bibfnamefont {B.~A.}\ \bibnamefont
  {Dobrescu}},\ }\href {\doibase 10.1103/PhysRevD.55.5826} {\bibfield
  {journal} {\bibinfo  {journal} {Phys. Rev. D}\ }\textbf {\bibinfo {volume}
  {55}},\ \bibinfo {pages} {5826} (\bibinfo {year} {1997})},\ \Eprint
  {http://arxiv.org/abs/hep-ph/9609221} {arXiv:hep-ph/9609221} \BibitemShut
  {NoStop}%
\bibitem [{\citenamefont {Banks}\ and\ \citenamefont
  {Seiberg}(2011)}]{Banks:2010zn}%
  \BibitemOpen
  \bibfield  {author} {\bibinfo {author} {\bibfnamefont {T.}~\bibnamefont
  {Banks}}\ and\ \bibinfo {author} {\bibfnamefont {N.}~\bibnamefont
  {Seiberg}},\ }\href {\doibase 10.1103/PhysRevD.83.084019} {\bibfield
  {journal} {\bibinfo  {journal} {Phys. Rev. D}\ }\textbf {\bibinfo {volume}
  {83}},\ \bibinfo {pages} {084019} (\bibinfo {year} {2011})},\ \Eprint
  {http://arxiv.org/abs/1011.5120} {arXiv:1011.5120 [hep-th]} \BibitemShut
  {NoStop}%
\bibitem [{\citenamefont {Preskill}\ \emph {et~al.}(1983)\citenamefont
  {Preskill}, \citenamefont {Wise},\ and\ \citenamefont
  {Wilczek}}]{Preskill:1982cy}%
  \BibitemOpen
  \bibfield  {author} {\bibinfo {author} {\bibfnamefont {J.}~\bibnamefont
  {Preskill}}, \bibinfo {author} {\bibfnamefont {M.~B.}\ \bibnamefont {Wise}},
  \ and\ \bibinfo {author} {\bibfnamefont {F.}~\bibnamefont {Wilczek}},\ }\href
  {\doibase 10.1016/0370-2693(83)90637-8} {\bibfield  {journal} {\bibinfo
  {journal} {Phys. Lett. B}\ }\textbf {\bibinfo {volume} {120}},\ \bibinfo
  {pages} {127} (\bibinfo {year} {1983})}\BibitemShut {NoStop}%
\bibitem [{\citenamefont {Abbott}\ and\ \citenamefont
  {Sikivie}(1983)}]{Abbott:1982af}%
  \BibitemOpen
  \bibfield  {author} {\bibinfo {author} {\bibfnamefont {L.}~\bibnamefont
  {Abbott}}\ and\ \bibinfo {author} {\bibfnamefont {P.}~\bibnamefont
  {Sikivie}},\ }\href {\doibase 10.1016/0370-2693(83)90638-X} {\bibfield
  {journal} {\bibinfo  {journal} {Phys. Lett. B}\ }\textbf {\bibinfo {volume}
  {120}},\ \bibinfo {pages} {133} (\bibinfo {year} {1983})}\BibitemShut
  {NoStop}%
\bibitem [{\citenamefont {Dine}\ and\ \citenamefont
  {Fischler}(1983)}]{Dine:1982ah}%
  \BibitemOpen
  \bibfield  {author} {\bibinfo {author} {\bibfnamefont {M.}~\bibnamefont
  {Dine}}\ and\ \bibinfo {author} {\bibfnamefont {W.}~\bibnamefont
  {Fischler}},\ }\href {\doibase 10.1016/0370-2693(83)90639-1} {\bibfield
  {journal} {\bibinfo  {journal} {Phys. Lett. B}\ }\textbf {\bibinfo {volume}
  {120}},\ \bibinfo {pages} {137} (\bibinfo {year} {1983})}\BibitemShut
  {NoStop}%
\bibitem [{\citenamefont {Mayle}\ \emph {et~al.}(1988)\citenamefont {Mayle},
  \citenamefont {Wilson}, \citenamefont {Ellis}, \citenamefont {Olive},
  \citenamefont {Schramm},\ and\ \citenamefont {Steigman}}]{Mayle:1987as}%
  \BibitemOpen
  \bibfield  {author} {\bibinfo {author} {\bibfnamefont {R.}~\bibnamefont
  {Mayle}}, \bibinfo {author} {\bibfnamefont {J.~R.}\ \bibnamefont {Wilson}},
  \bibinfo {author} {\bibfnamefont {J.~R.}\ \bibnamefont {Ellis}}, \bibinfo
  {author} {\bibfnamefont {K.~A.}\ \bibnamefont {Olive}}, \bibinfo {author}
  {\bibfnamefont {D.~N.}\ \bibnamefont {Schramm}}, \ and\ \bibinfo {author}
  {\bibfnamefont {G.}~\bibnamefont {Steigman}},\ }\href {\doibase
  10.1016/0370-2693(88)91595-X} {\bibfield  {journal} {\bibinfo  {journal}
  {Phys. Lett. B}\ }\textbf {\bibinfo {volume} {203}},\ \bibinfo {pages} {188}
  (\bibinfo {year} {1988})}\BibitemShut {NoStop}%
\bibitem [{\citenamefont {Raffelt}\ and\ \citenamefont
  {Seckel}(1988)}]{Raffelt:1987yt}%
  \BibitemOpen
  \bibfield  {author} {\bibinfo {author} {\bibfnamefont {G.}~\bibnamefont
  {Raffelt}}\ and\ \bibinfo {author} {\bibfnamefont {D.}~\bibnamefont
  {Seckel}},\ }\href {\doibase 10.1103/PhysRevLett.60.1793} {\bibfield
  {journal} {\bibinfo  {journal} {Phys. Rev. Lett.}\ }\textbf {\bibinfo
  {volume} {60}},\ \bibinfo {pages} {1793} (\bibinfo {year}
  {1988})}\BibitemShut {NoStop}%
\bibitem [{\citenamefont {Chun}\ and\ \citenamefont
  {Lukas}(1992)}]{Chun:1992bn}%
  \BibitemOpen
  \bibfield  {author} {\bibinfo {author} {\bibfnamefont {E.}~\bibnamefont
  {Chun}}\ and\ \bibinfo {author} {\bibfnamefont {A.}~\bibnamefont {Lukas}},\
  }\href {\doibase 10.1016/0370-2693(92)91266-C} {\bibfield  {journal}
  {\bibinfo  {journal} {Phys. Lett. B}\ }\textbf {\bibinfo {volume} {297}},\
  \bibinfo {pages} {298} (\bibinfo {year} {1992})},\ \Eprint
  {http://arxiv.org/abs/hep-ph/9209208} {arXiv:hep-ph/9209208} \BibitemShut
  {NoStop}%
\bibitem [{\citenamefont {Bastero-Gil}\ and\ \citenamefont
  {King}(1998)}]{BasteroGil:1997vn}%
  \BibitemOpen
  \bibfield  {author} {\bibinfo {author} {\bibfnamefont {M.}~\bibnamefont
  {Bastero-Gil}}\ and\ \bibinfo {author} {\bibfnamefont {S.}~\bibnamefont
  {King}},\ }\href {\doibase 10.1016/S0370-2693(98)00124-5} {\bibfield
  {journal} {\bibinfo  {journal} {Phys. Lett. B}\ }\textbf {\bibinfo {volume}
  {423}},\ \bibinfo {pages} {27} (\bibinfo {year} {1998})},\ \Eprint
  {http://arxiv.org/abs/hep-ph/9709502} {arXiv:hep-ph/9709502} \BibitemShut
  {NoStop}%
\bibitem [{\citenamefont {Babu}\ \emph {et~al.}(2003)\citenamefont {Babu},
  \citenamefont {Gogoladze},\ and\ \citenamefont {Wang}}]{Babu:2002ic}%
  \BibitemOpen
  \bibfield  {author} {\bibinfo {author} {\bibfnamefont {K.}~\bibnamefont
  {Babu}}, \bibinfo {author} {\bibfnamefont {I.}~\bibnamefont {Gogoladze}}, \
  and\ \bibinfo {author} {\bibfnamefont {K.}~\bibnamefont {Wang}},\ }\href
  {\doibase 10.1016/S0370-2693(03)00411-8} {\bibfield  {journal} {\bibinfo
  {journal} {Phys. Lett. B}\ }\textbf {\bibinfo {volume} {560}},\ \bibinfo
  {pages} {214} (\bibinfo {year} {2003})},\ \Eprint
  {http://arxiv.org/abs/hep-ph/0212339} {arXiv:hep-ph/0212339} \BibitemShut
  {NoStop}%
\bibitem [{\citenamefont {Dias}\ \emph {et~al.}(2004)\citenamefont {Dias},
  \citenamefont {Pleitez},\ and\ \citenamefont {Tonasse}}]{Dias:2002hz}%
  \BibitemOpen
  \bibfield  {author} {\bibinfo {author} {\bibfnamefont {A.~G.}\ \bibnamefont
  {Dias}}, \bibinfo {author} {\bibfnamefont {V.}~\bibnamefont {Pleitez}}, \
  and\ \bibinfo {author} {\bibfnamefont {M.}~\bibnamefont {Tonasse}},\ }\href
  {\doibase 10.1103/PhysRevD.69.015007} {\bibfield  {journal} {\bibinfo
  {journal} {Phys. Rev. D}\ }\textbf {\bibinfo {volume} {69}},\ \bibinfo
  {pages} {015007} (\bibinfo {year} {2004})},\ \Eprint
  {http://arxiv.org/abs/hep-ph/0210172} {arXiv:hep-ph/0210172} \BibitemShut
  {NoStop}%
\bibitem [{\citenamefont {Harigaya}\ \emph {et~al.}(2013)\citenamefont
  {Harigaya}, \citenamefont {Ibe}, \citenamefont {Schmitz},\ and\ \citenamefont
  {Yanagida}}]{Harigaya:2013vja}%
  \BibitemOpen
  \bibfield  {author} {\bibinfo {author} {\bibfnamefont {K.}~\bibnamefont
  {Harigaya}}, \bibinfo {author} {\bibfnamefont {M.}~\bibnamefont {Ibe}},
  \bibinfo {author} {\bibfnamefont {K.}~\bibnamefont {Schmitz}}, \ and\
  \bibinfo {author} {\bibfnamefont {T.~T.}\ \bibnamefont {Yanagida}},\ }\href
  {\doibase 10.1103/PhysRevD.88.075022} {\bibfield  {journal} {\bibinfo
  {journal} {Phys. Rev. D}\ }\textbf {\bibinfo {volume} {88}},\ \bibinfo
  {pages} {075022} (\bibinfo {year} {2013})},\ \Eprint
  {http://arxiv.org/abs/1308.1227} {arXiv:1308.1227 [hep-ph]} \BibitemShut
  {NoStop}%
\bibitem [{\citenamefont {Harigaya}\ \emph {et~al.}(2015)\citenamefont
  {Harigaya}, \citenamefont {Ibe}, \citenamefont {Schmitz},\ and\ \citenamefont
  {Yanagida}}]{Harigaya:2015soa}%
  \BibitemOpen
  \bibfield  {author} {\bibinfo {author} {\bibfnamefont {K.}~\bibnamefont
  {Harigaya}}, \bibinfo {author} {\bibfnamefont {M.}~\bibnamefont {Ibe}},
  \bibinfo {author} {\bibfnamefont {K.}~\bibnamefont {Schmitz}}, \ and\
  \bibinfo {author} {\bibfnamefont {T.~T.}\ \bibnamefont {Yanagida}},\ }\href
  {\doibase 10.1103/PhysRevD.92.075003} {\bibfield  {journal} {\bibinfo
  {journal} {Phys. Rev. D}\ }\textbf {\bibinfo {volume} {92}},\ \bibinfo
  {pages} {075003} (\bibinfo {year} {2015})},\ \Eprint
  {http://arxiv.org/abs/1505.07388} {arXiv:1505.07388 [hep-ph]} \BibitemShut
  {NoStop}%
\bibitem [{\citenamefont {Fukuda}\ \emph {et~al.}(2017)\citenamefont {Fukuda},
  \citenamefont {Ibe}, \citenamefont {Suzuki},\ and\ \citenamefont
  {Yanagida}}]{Fukuda:2017ylt}%
  \BibitemOpen
  \bibfield  {author} {\bibinfo {author} {\bibfnamefont {H.}~\bibnamefont
  {Fukuda}}, \bibinfo {author} {\bibfnamefont {M.}~\bibnamefont {Ibe}},
  \bibinfo {author} {\bibfnamefont {M.}~\bibnamefont {Suzuki}}, \ and\ \bibinfo
  {author} {\bibfnamefont {T.~T.}\ \bibnamefont {Yanagida}},\ }\href {\doibase
  10.1016/j.physletb.2017.05.071} {\bibfield  {journal} {\bibinfo  {journal}
  {Phys. Lett. B}\ }\textbf {\bibinfo {volume} {771}},\ \bibinfo {pages} {327}
  (\bibinfo {year} {2017})},\ \Eprint {http://arxiv.org/abs/1703.01112}
  {arXiv:1703.01112 [hep-ph]} \BibitemShut {NoStop}%
\bibitem [{\citenamefont {Duerr}\ \emph {et~al.}(2018)\citenamefont {Duerr},
  \citenamefont {Schmidt-Hoberg},\ and\ \citenamefont {Unwin}}]{Duerr:2017amf}%
  \BibitemOpen
  \bibfield  {author} {\bibinfo {author} {\bibfnamefont {M.}~\bibnamefont
  {Duerr}}, \bibinfo {author} {\bibfnamefont {K.}~\bibnamefont
  {Schmidt-Hoberg}}, \ and\ \bibinfo {author} {\bibfnamefont {J.}~\bibnamefont
  {Unwin}},\ }\href {\doibase 10.1016/j.physletb.2018.03.054} {\bibfield
  {journal} {\bibinfo  {journal} {Phys. Lett. B}\ }\textbf {\bibinfo {volume}
  {780}},\ \bibinfo {pages} {553} (\bibinfo {year} {2018})},\ \Eprint
  {http://arxiv.org/abs/1712.01841} {arXiv:1712.01841 [hep-ph]} \BibitemShut
  {NoStop}%
\bibitem [{\citenamefont {Fukuda}\ \emph {et~al.}(2018)\citenamefont {Fukuda},
  \citenamefont {Ibe}, \citenamefont {Suzuki},\ and\ \citenamefont
  {Yanagida}}]{Fukuda:2018oco}%
  \BibitemOpen
  \bibfield  {author} {\bibinfo {author} {\bibfnamefont {H.}~\bibnamefont
  {Fukuda}}, \bibinfo {author} {\bibfnamefont {M.}~\bibnamefont {Ibe}},
  \bibinfo {author} {\bibfnamefont {M.}~\bibnamefont {Suzuki}}, \ and\ \bibinfo
  {author} {\bibfnamefont {T.~T.}\ \bibnamefont {Yanagida}},\ }\href {\doibase
  10.1007/JHEP07(2018)128} {\bibfield  {journal} {\bibinfo  {journal} {JHEP}\
  }\textbf {\bibinfo {volume} {07}},\ \bibinfo {pages} {128} (\bibinfo {year}
  {2018})},\ \Eprint {http://arxiv.org/abs/1803.00759} {arXiv:1803.00759
  [hep-ph]} \BibitemShut {NoStop}%
\bibitem [{\citenamefont {Bonnefoy}\ \emph {et~al.}(2019)\citenamefont
  {Bonnefoy}, \citenamefont {Dudas},\ and\ \citenamefont
  {Pokorski}}]{Bonnefoy:2018ibr}%
  \BibitemOpen
  \bibfield  {author} {\bibinfo {author} {\bibfnamefont {Q.}~\bibnamefont
  {Bonnefoy}}, \bibinfo {author} {\bibfnamefont {E.}~\bibnamefont {Dudas}}, \
  and\ \bibinfo {author} {\bibfnamefont {S.}~\bibnamefont {Pokorski}},\ }\href
  {\doibase 10.1140/epjc/s10052-018-6528-z} {\bibfield  {journal} {\bibinfo
  {journal} {Eur. Phys. J. C}\ }\textbf {\bibinfo {volume} {79}},\ \bibinfo
  {pages} {31} (\bibinfo {year} {2019})},\ \Eprint
  {http://arxiv.org/abs/1804.01112} {arXiv:1804.01112 [hep-ph]} \BibitemShut
  {NoStop}%
\bibitem [{\citenamefont {Ibe}\ \emph {et~al.}(2018)\citenamefont {Ibe},
  \citenamefont {Suzuki},\ and\ \citenamefont {Yanagida}}]{Ibe:2018hir}%
  \BibitemOpen
  \bibfield  {author} {\bibinfo {author} {\bibfnamefont {M.}~\bibnamefont
  {Ibe}}, \bibinfo {author} {\bibfnamefont {M.}~\bibnamefont {Suzuki}}, \ and\
  \bibinfo {author} {\bibfnamefont {T.~T.}\ \bibnamefont {Yanagida}},\ }\href
  {\doibase 10.1007/JHEP08(2018)049} {\bibfield  {journal} {\bibinfo  {journal}
  {JHEP}\ }\textbf {\bibinfo {volume} {08}},\ \bibinfo {pages} {049} (\bibinfo
  {year} {2018})},\ \Eprint {http://arxiv.org/abs/1805.10029} {arXiv:1805.10029
  [hep-ph]} \BibitemShut {NoStop}%
\bibitem [{\citenamefont {Randall}(1992)}]{Randall:1992ut}%
  \BibitemOpen
  \bibfield  {author} {\bibinfo {author} {\bibfnamefont {L.}~\bibnamefont
  {Randall}},\ }\href {\doibase 10.1016/0370-2693(92)91928-3} {\bibfield
  {journal} {\bibinfo  {journal} {Phys. Lett. B}\ }\textbf {\bibinfo {volume}
  {284}},\ \bibinfo {pages} {77} (\bibinfo {year} {1992})}\BibitemShut
  {NoStop}%
\bibitem [{\citenamefont {Redi}\ and\ \citenamefont
  {Sato}(2016)}]{Redi:2016esr}%
  \BibitemOpen
  \bibfield  {author} {\bibinfo {author} {\bibfnamefont {M.}~\bibnamefont
  {Redi}}\ and\ \bibinfo {author} {\bibfnamefont {R.}~\bibnamefont {Sato}},\
  }\href {\doibase 10.1007/JHEP05(2016)104} {\bibfield  {journal} {\bibinfo
  {journal} {JHEP}\ }\textbf {\bibinfo {volume} {05}},\ \bibinfo {pages} {104}
  (\bibinfo {year} {2016})},\ \Eprint {http://arxiv.org/abs/1602.05427}
  {arXiv:1602.05427 [hep-ph]} \BibitemShut {NoStop}%
\bibitem [{\citenamefont {Di~Luzio}\ \emph {et~al.}(2017)\citenamefont
  {Di~Luzio}, \citenamefont {Nardi},\ and\ \citenamefont
  {Ubaldi}}]{DiLuzio:2017tjx}%
  \BibitemOpen
  \bibfield  {author} {\bibinfo {author} {\bibfnamefont {L.}~\bibnamefont
  {Di~Luzio}}, \bibinfo {author} {\bibfnamefont {E.}~\bibnamefont {Nardi}}, \
  and\ \bibinfo {author} {\bibfnamefont {L.}~\bibnamefont {Ubaldi}},\ }\href
  {\doibase 10.1103/PhysRevLett.119.011801} {\bibfield  {journal} {\bibinfo
  {journal} {Phys. Rev. Lett.}\ }\textbf {\bibinfo {volume} {119}},\ \bibinfo
  {pages} {011801} (\bibinfo {year} {2017})},\ \Eprint
  {http://arxiv.org/abs/1704.01122} {arXiv:1704.01122 [hep-ph]} \BibitemShut
  {NoStop}%
\bibitem [{\citenamefont {Lillard}\ and\ \citenamefont
  {Tait}(2018)}]{Lillard:2018fdt}%
  \BibitemOpen
  \bibfield  {author} {\bibinfo {author} {\bibfnamefont {B.}~\bibnamefont
  {Lillard}}\ and\ \bibinfo {author} {\bibfnamefont {T.~M.}\ \bibnamefont
  {Tait}},\ }\href {\doibase 10.1007/JHEP11(2018)199} {\bibfield  {journal}
  {\bibinfo  {journal} {JHEP}\ }\textbf {\bibinfo {volume} {11}},\ \bibinfo
  {pages} {199} (\bibinfo {year} {2018})},\ \Eprint
  {http://arxiv.org/abs/1811.03089} {arXiv:1811.03089 [hep-ph]} \BibitemShut
  {NoStop}%
\bibitem [{\citenamefont {Lee}\ and\ \citenamefont {Yin}(2019)}]{Lee:2018yak}%
  \BibitemOpen
  \bibfield  {author} {\bibinfo {author} {\bibfnamefont {H.-S.}\ \bibnamefont
  {Lee}}\ and\ \bibinfo {author} {\bibfnamefont {W.}~\bibnamefont {Yin}},\
  }\href {\doibase 10.1103/PhysRevD.99.015041} {\bibfield  {journal} {\bibinfo
  {journal} {Phys. Rev. D}\ }\textbf {\bibinfo {volume} {99}},\ \bibinfo
  {pages} {015041} (\bibinfo {year} {2019})},\ \Eprint
  {http://arxiv.org/abs/1811.04039} {arXiv:1811.04039 [hep-ph]} \BibitemShut
  {NoStop}%
\bibitem [{\citenamefont {Gavela}\ \emph {et~al.}(2019)\citenamefont {Gavela},
  \citenamefont {Ibe}, \citenamefont {Quilez},\ and\ \citenamefont
  {Yanagida}}]{Gavela:2018paw}%
  \BibitemOpen
  \bibfield  {author} {\bibinfo {author} {\bibfnamefont {M.}~\bibnamefont
  {Gavela}}, \bibinfo {author} {\bibfnamefont {M.}~\bibnamefont {Ibe}},
  \bibinfo {author} {\bibfnamefont {P.}~\bibnamefont {Quilez}}, \ and\ \bibinfo
  {author} {\bibfnamefont {T.}~\bibnamefont {Yanagida}},\ }\href {\doibase
  10.1140/epjc/s10052-019-7046-3} {\bibfield  {journal} {\bibinfo  {journal}
  {Eur. Phys. J. C}\ }\textbf {\bibinfo {volume} {79}},\ \bibinfo {pages} {542}
  (\bibinfo {year} {2019})},\ \Eprint {http://arxiv.org/abs/1812.08174}
  {arXiv:1812.08174 [hep-ph]} \BibitemShut {NoStop}%
\bibitem [{\citenamefont {Buttazzo}\ \emph {et~al.}(2020)\citenamefont
  {Buttazzo}, \citenamefont {Di~Luzio}, \citenamefont {Ghorbani}, \citenamefont
  {Gross}, \citenamefont {Landini}, \citenamefont {Strumia}, \citenamefont
  {Teresi},\ and\ \citenamefont {Wang}}]{Buttazzo:2019mvl}%
  \BibitemOpen
  \bibfield  {author} {\bibinfo {author} {\bibfnamefont {D.}~\bibnamefont
  {Buttazzo}}, \bibinfo {author} {\bibfnamefont {L.}~\bibnamefont {Di~Luzio}},
  \bibinfo {author} {\bibfnamefont {P.}~\bibnamefont {Ghorbani}}, \bibinfo
  {author} {\bibfnamefont {C.}~\bibnamefont {Gross}}, \bibinfo {author}
  {\bibfnamefont {G.}~\bibnamefont {Landini}}, \bibinfo {author} {\bibfnamefont
  {A.}~\bibnamefont {Strumia}}, \bibinfo {author} {\bibfnamefont
  {D.}~\bibnamefont {Teresi}}, \ and\ \bibinfo {author} {\bibfnamefont {J.-W.}\
  \bibnamefont {Wang}},\ }\href {\doibase 10.1007/JHEP01(2020)130} {\bibfield
  {journal} {\bibinfo  {journal} {JHEP}\ }\textbf {\bibinfo {volume} {01}},\
  \bibinfo {pages} {130} (\bibinfo {year} {2020})},\ \Eprint
  {http://arxiv.org/abs/1911.04502} {arXiv:1911.04502 [hep-ph]} \BibitemShut
  {NoStop}%
\bibitem [{\citenamefont {Ardu}\ \emph {et~al.}(2020)\citenamefont {Ardu},
  \citenamefont {Di~Luzio}, \citenamefont {Landini}, \citenamefont {Strumia},
  \citenamefont {Teresi},\ and\ \citenamefont {Wang}}]{Ardu:2020qmo}%
  \BibitemOpen
  \bibfield  {author} {\bibinfo {author} {\bibfnamefont {M.}~\bibnamefont
  {Ardu}}, \bibinfo {author} {\bibfnamefont {L.}~\bibnamefont {Di~Luzio}},
  \bibinfo {author} {\bibfnamefont {G.}~\bibnamefont {Landini}}, \bibinfo
  {author} {\bibfnamefont {A.}~\bibnamefont {Strumia}}, \bibinfo {author}
  {\bibfnamefont {D.}~\bibnamefont {Teresi}}, \ and\ \bibinfo {author}
  {\bibfnamefont {J.-W.}\ \bibnamefont {Wang}},\ }\href {\doibase
  10.1007/JHEP11(2020)090} {\bibfield  {journal} {\bibinfo  {journal} {JHEP}\
  }\textbf {\bibinfo {volume} {11}},\ \bibinfo {pages} {090} (\bibinfo {year}
  {2020})},\ \Eprint {http://arxiv.org/abs/2007.12663} {arXiv:2007.12663
  [hep-ph]} \BibitemShut {NoStop}%
\bibitem [{\citenamefont {Yin}(2020)}]{Yin:2020dfn}%
  \BibitemOpen
  \bibfield  {author} {\bibinfo {author} {\bibfnamefont {W.}~\bibnamefont
  {Yin}},\ }\href {\doibase 10.1007/JHEP10(2020)032} {\bibfield  {journal}
  {\bibinfo  {journal} {JHEP}\ }\textbf {\bibinfo {volume} {10}},\ \bibinfo
  {pages} {032} (\bibinfo {year} {2020})},\ \Eprint
  {http://arxiv.org/abs/2007.13320} {arXiv:2007.13320 [hep-ph]} \BibitemShut
  {NoStop}%
\bibitem [{\citenamefont {Cheng}\ and\ \citenamefont
  {Kaplan}(2001)}]{Cheng:2001ys}%
  \BibitemOpen
  \bibfield  {author} {\bibinfo {author} {\bibfnamefont {H.-C.}\ \bibnamefont
  {Cheng}}\ and\ \bibinfo {author} {\bibfnamefont {D.~E.}\ \bibnamefont
  {Kaplan}},\ }\href@noop {} {\  (\bibinfo {year} {2001})},\ \Eprint
  {http://arxiv.org/abs/hep-ph/0103346} {arXiv:hep-ph/0103346} \BibitemShut
  {NoStop}%
\bibitem [{\citenamefont {Izawa}\ \emph {et~al.}(2002)\citenamefont {Izawa},
  \citenamefont {Watari},\ and\ \citenamefont {Yanagida}}]{Izawa:2002qk}%
  \BibitemOpen
  \bibfield  {author} {\bibinfo {author} {\bibfnamefont {K.}~\bibnamefont
  {Izawa}}, \bibinfo {author} {\bibfnamefont {T.}~\bibnamefont {Watari}}, \
  and\ \bibinfo {author} {\bibfnamefont {T.}~\bibnamefont {Yanagida}},\ }\href
  {\doibase 10.1016/S0370-2693(02)01663-5} {\bibfield  {journal} {\bibinfo
  {journal} {Phys. Lett. B}\ }\textbf {\bibinfo {volume} {534}},\ \bibinfo
  {pages} {93} (\bibinfo {year} {2002})},\ \Eprint
  {http://arxiv.org/abs/hep-ph/0202171} {arXiv:hep-ph/0202171} \BibitemShut
  {NoStop}%
\bibitem [{\citenamefont {Fukunaga}\ and\ \citenamefont
  {Izawa}(2003)}]{Fukunaga:2003sz}%
  \BibitemOpen
  \bibfield  {author} {\bibinfo {author} {\bibfnamefont {A.}~\bibnamefont
  {Fukunaga}}\ and\ \bibinfo {author} {\bibfnamefont {K.}~\bibnamefont
  {Izawa}},\ }\href {\doibase 10.1016/S0370-2693(03)00584-7} {\bibfield
  {journal} {\bibinfo  {journal} {Phys. Lett. B}\ }\textbf {\bibinfo {volume}
  {562}},\ \bibinfo {pages} {251} (\bibinfo {year} {2003})},\ \Eprint
  {http://arxiv.org/abs/hep-ph/0301273} {arXiv:hep-ph/0301273} \BibitemShut
  {NoStop}%
\bibitem [{\citenamefont {Choi}(2004)}]{Choi:2003wr}%
  \BibitemOpen
  \bibfield  {author} {\bibinfo {author} {\bibfnamefont {K.-w.}\ \bibnamefont
  {Choi}},\ }\href {\doibase 10.1103/PhysRevLett.92.101602} {\bibfield
  {journal} {\bibinfo  {journal} {Phys. Rev. Lett.}\ }\textbf {\bibinfo
  {volume} {92}},\ \bibinfo {pages} {101602} (\bibinfo {year} {2004})},\
  \Eprint {http://arxiv.org/abs/hep-ph/0308024} {arXiv:hep-ph/0308024}
  \BibitemShut {NoStop}%
\bibitem [{\citenamefont {Izawa}\ \emph {et~al.}(2004)\citenamefont {Izawa},
  \citenamefont {Watari},\ and\ \citenamefont {Yanagida}}]{Izawa:2004bi}%
  \BibitemOpen
  \bibfield  {author} {\bibinfo {author} {\bibfnamefont {K.}~\bibnamefont
  {Izawa}}, \bibinfo {author} {\bibfnamefont {T.}~\bibnamefont {Watari}}, \
  and\ \bibinfo {author} {\bibfnamefont {T.}~\bibnamefont {Yanagida}},\ }\href
  {\doibase 10.1016/j.physletb.2004.03.061} {\bibfield  {journal} {\bibinfo
  {journal} {Phys. Lett. B}\ }\textbf {\bibinfo {volume} {589}},\ \bibinfo
  {pages} {141} (\bibinfo {year} {2004})},\ \Eprint
  {http://arxiv.org/abs/hep-ph/0403090} {arXiv:hep-ph/0403090} \BibitemShut
  {NoStop}%
\bibitem [{\citenamefont {Flacke}\ \emph {et~al.}(2007)\citenamefont {Flacke},
  \citenamefont {Gripaios}, \citenamefont {March-Russell},\ and\ \citenamefont
  {Maybury}}]{Flacke:2006ad}%
  \BibitemOpen
  \bibfield  {author} {\bibinfo {author} {\bibfnamefont {T.}~\bibnamefont
  {Flacke}}, \bibinfo {author} {\bibfnamefont {B.}~\bibnamefont {Gripaios}},
  \bibinfo {author} {\bibfnamefont {J.}~\bibnamefont {March-Russell}}, \ and\
  \bibinfo {author} {\bibfnamefont {D.}~\bibnamefont {Maybury}},\ }\href
  {\doibase 10.1088/1126-6708/2007/01/061} {\bibfield  {journal} {\bibinfo
  {journal} {JHEP}\ }\textbf {\bibinfo {volume} {01}},\ \bibinfo {pages} {061}
  (\bibinfo {year} {2007})},\ \Eprint {http://arxiv.org/abs/hep-ph/0611278}
  {arXiv:hep-ph/0611278} \BibitemShut {NoStop}%
\bibitem [{\citenamefont {Kawasaki}\ \emph
  {et~al.}(2015{\natexlab{a}})\citenamefont {Kawasaki}, \citenamefont
  {Yamada},\ and\ \citenamefont {Yanagida}}]{Kawasaki:2015lea}%
  \BibitemOpen
  \bibfield  {author} {\bibinfo {author} {\bibfnamefont {M.}~\bibnamefont
  {Kawasaki}}, \bibinfo {author} {\bibfnamefont {M.}~\bibnamefont {Yamada}}, \
  and\ \bibinfo {author} {\bibfnamefont {T.~T.}\ \bibnamefont {Yanagida}},\
  }\href {\doibase 10.1016/j.physletb.2015.08.043} {\bibfield  {journal}
  {\bibinfo  {journal} {Phys. Lett. B}\ }\textbf {\bibinfo {volume} {750}},\
  \bibinfo {pages} {12} (\bibinfo {year} {2015}{\natexlab{a}})},\ \Eprint
  {http://arxiv.org/abs/1506.05214} {arXiv:1506.05214 [hep-ph]} \BibitemShut
  {NoStop}%
\bibitem [{\citenamefont {Yamada}\ \emph {et~al.}(2016)\citenamefont {Yamada},
  \citenamefont {Yanagida},\ and\ \citenamefont {Yonekura}}]{Yamada:2015waa}%
  \BibitemOpen
  \bibfield  {author} {\bibinfo {author} {\bibfnamefont {M.}~\bibnamefont
  {Yamada}}, \bibinfo {author} {\bibfnamefont {T.~T.}\ \bibnamefont
  {Yanagida}}, \ and\ \bibinfo {author} {\bibfnamefont {K.}~\bibnamefont
  {Yonekura}},\ }\href {\doibase 10.1103/PhysRevLett.116.051801} {\bibfield
  {journal} {\bibinfo  {journal} {Phys. Rev. Lett.}\ }\textbf {\bibinfo
  {volume} {116}},\ \bibinfo {pages} {051801} (\bibinfo {year} {2016})},\
  \Eprint {http://arxiv.org/abs/1510.06504} {arXiv:1510.06504 [hep-ph]}
  \BibitemShut {NoStop}%
\bibitem [{\citenamefont {Cox}\ \emph {et~al.}(2020)\citenamefont {Cox},
  \citenamefont {Gherghetta},\ and\ \citenamefont {Nguyen}}]{Cox:2019rro}%
  \BibitemOpen
  \bibfield  {author} {\bibinfo {author} {\bibfnamefont {P.}~\bibnamefont
  {Cox}}, \bibinfo {author} {\bibfnamefont {T.}~\bibnamefont {Gherghetta}}, \
  and\ \bibinfo {author} {\bibfnamefont {M.~D.}\ \bibnamefont {Nguyen}},\
  }\href {\doibase 10.1007/JHEP01(2020)188} {\bibfield  {journal} {\bibinfo
  {journal} {JHEP}\ }\textbf {\bibinfo {volume} {01}},\ \bibinfo {pages} {188}
  (\bibinfo {year} {2020})},\ \Eprint {http://arxiv.org/abs/1911.09385}
  {arXiv:1911.09385 [hep-ph]} \BibitemShut {NoStop}%
\bibitem [{\citenamefont {Kim}(1979)}]{Kim:1979if}%
  \BibitemOpen
  \bibfield  {author} {\bibinfo {author} {\bibfnamefont {J.~E.}\ \bibnamefont
  {Kim}},\ }\href {\doibase 10.1103/PhysRevLett.43.103} {\bibfield  {journal}
  {\bibinfo  {journal} {Phys. Rev. Lett.}\ }\textbf {\bibinfo {volume} {43}},\
  \bibinfo {pages} {103} (\bibinfo {year} {1979})}\BibitemShut {NoStop}%
\bibitem [{\citenamefont {Fujikura}\ \emph {et~al.}(2020)\citenamefont
  {Fujikura}, \citenamefont {Nakai},\ and\ \citenamefont
  {Yamada}}]{Fujikura:2019oyi}%
  \BibitemOpen
  \bibfield  {author} {\bibinfo {author} {\bibfnamefont {K.}~\bibnamefont
  {Fujikura}}, \bibinfo {author} {\bibfnamefont {Y.}~\bibnamefont {Nakai}}, \
  and\ \bibinfo {author} {\bibfnamefont {M.}~\bibnamefont {Yamada}},\ }\href
  {\doibase 10.1007/JHEP02(2020)111} {\bibfield  {journal} {\bibinfo  {journal}
  {JHEP}\ }\textbf {\bibinfo {volume} {02}},\ \bibinfo {pages} {111} (\bibinfo
  {year} {2020})},\ \Eprint {http://arxiv.org/abs/1910.07546} {arXiv:1910.07546
  [hep-ph]} \BibitemShut {NoStop}%
\bibitem [{\citenamefont {von Harling}\ and\ \citenamefont
  {Servant}(2018)}]{vonHarling:2017yew}%
  \BibitemOpen
  \bibfield  {author} {\bibinfo {author} {\bibfnamefont {B.}~\bibnamefont {von
  Harling}}\ and\ \bibinfo {author} {\bibfnamefont {G.}~\bibnamefont
  {Servant}},\ }\href {\doibase 10.1007/JHEP01(2018)159} {\bibfield  {journal}
  {\bibinfo  {journal} {JHEP}\ }\textbf {\bibinfo {volume} {01}},\ \bibinfo
  {pages} {159} (\bibinfo {year} {2018})},\ \Eprint
  {http://arxiv.org/abs/1711.11554} {arXiv:1711.11554 [hep-ph]} \BibitemShut
  {NoStop}%
\bibitem [{\citenamefont {Randall}\ and\ \citenamefont
  {Sundrum}(1999)}]{Randall:1999ee}%
  \BibitemOpen
  \bibfield  {author} {\bibinfo {author} {\bibfnamefont {L.}~\bibnamefont
  {Randall}}\ and\ \bibinfo {author} {\bibfnamefont {R.}~\bibnamefont
  {Sundrum}},\ }\href {\doibase 10.1103/PhysRevLett.83.3370} {\bibfield
  {journal} {\bibinfo  {journal} {Phys. Rev. Lett.}\ }\textbf {\bibinfo
  {volume} {83}},\ \bibinfo {pages} {3370} (\bibinfo {year} {1999})},\ \Eprint
  {http://arxiv.org/abs/hep-ph/9905221} {arXiv:hep-ph/9905221} \BibitemShut
  {NoStop}%
\bibitem [{\citenamefont {Agashe}\ \emph {et~al.}(2002)\citenamefont {Agashe},
  \citenamefont {Delgado},\ and\ \citenamefont {Sundrum}}]{Agashe:2002bx}%
  \BibitemOpen
  \bibfield  {author} {\bibinfo {author} {\bibfnamefont {K.}~\bibnamefont
  {Agashe}}, \bibinfo {author} {\bibfnamefont {A.}~\bibnamefont {Delgado}}, \
  and\ \bibinfo {author} {\bibfnamefont {R.}~\bibnamefont {Sundrum}},\ }\href
  {\doibase 10.1016/S0550-3213(02)00740-X} {\bibfield  {journal} {\bibinfo
  {journal} {Nucl. Phys. B}\ }\textbf {\bibinfo {volume} {643}},\ \bibinfo
  {pages} {172} (\bibinfo {year} {2002})},\ \Eprint
  {http://arxiv.org/abs/hep-ph/0206099} {arXiv:hep-ph/0206099} \BibitemShut
  {NoStop}%
\bibitem [{\citenamefont {Csaki}\ \emph {et~al.}(2007)\citenamefont {Csaki},
  \citenamefont {Hubisz},\ and\ \citenamefont {Lee}}]{Csaki:2007ns}%
  \BibitemOpen
  \bibfield  {author} {\bibinfo {author} {\bibfnamefont {C.}~\bibnamefont
  {Csaki}}, \bibinfo {author} {\bibfnamefont {J.}~\bibnamefont {Hubisz}}, \
  and\ \bibinfo {author} {\bibfnamefont {S.~J.}\ \bibnamefont {Lee}},\ }\href
  {\doibase 10.1103/PhysRevD.76.125015} {\bibfield  {journal} {\bibinfo
  {journal} {Phys. Rev. D}\ }\textbf {\bibinfo {volume} {76}},\ \bibinfo
  {pages} {125015} (\bibinfo {year} {2007})},\ \Eprint
  {http://arxiv.org/abs/0705.3844} {arXiv:0705.3844 [hep-ph]} \BibitemShut
  {NoStop}%
\bibitem [{\citenamefont {Agashe}\ \emph {et~al.}(2007)\citenamefont {Agashe},
  \citenamefont {Davoudiasl}, \citenamefont {Perez},\ and\ \citenamefont
  {Soni}}]{Agashe:2007zd}%
  \BibitemOpen
  \bibfield  {author} {\bibinfo {author} {\bibfnamefont {K.}~\bibnamefont
  {Agashe}}, \bibinfo {author} {\bibfnamefont {H.}~\bibnamefont {Davoudiasl}},
  \bibinfo {author} {\bibfnamefont {G.}~\bibnamefont {Perez}}, \ and\ \bibinfo
  {author} {\bibfnamefont {A.}~\bibnamefont {Soni}},\ }\href {\doibase
  10.1103/PhysRevD.76.036006} {\bibfield  {journal} {\bibinfo  {journal} {Phys.
  Rev. D}\ }\textbf {\bibinfo {volume} {76}},\ \bibinfo {pages} {036006}
  (\bibinfo {year} {2007})},\ \Eprint {http://arxiv.org/abs/hep-ph/0701186}
  {arXiv:hep-ph/0701186} \BibitemShut {NoStop}%
\bibitem [{\citenamefont {Wantz}\ and\ \citenamefont
  {Shellard}(2010)}]{Wantz:2009it}%
  \BibitemOpen
  \bibfield  {author} {\bibinfo {author} {\bibfnamefont {O.}~\bibnamefont
  {Wantz}}\ and\ \bibinfo {author} {\bibfnamefont {E.}~\bibnamefont
  {Shellard}},\ }\href {\doibase 10.1103/PhysRevD.82.123508} {\bibfield
  {journal} {\bibinfo  {journal} {Phys. Rev. D}\ }\textbf {\bibinfo {volume}
  {82}},\ \bibinfo {pages} {123508} (\bibinfo {year} {2010})},\ \Eprint
  {http://arxiv.org/abs/0910.1066} {arXiv:0910.1066 [astro-ph.CO]} \BibitemShut
  {NoStop}%
\bibitem [{\citenamefont {Kawasaki}\ \emph
  {et~al.}(2015{\natexlab{b}})\citenamefont {Kawasaki}, \citenamefont
  {Saikawa},\ and\ \citenamefont {Sekiguchi}}]{Kawasaki:2014sqa}%
  \BibitemOpen
  \bibfield  {author} {\bibinfo {author} {\bibfnamefont {M.}~\bibnamefont
  {Kawasaki}}, \bibinfo {author} {\bibfnamefont {K.}~\bibnamefont {Saikawa}}, \
  and\ \bibinfo {author} {\bibfnamefont {T.}~\bibnamefont {Sekiguchi}},\ }\href
  {\doibase 10.1103/PhysRevD.91.065014} {\bibfield  {journal} {\bibinfo
  {journal} {Phys. Rev. D}\ }\textbf {\bibinfo {volume} {91}},\ \bibinfo
  {pages} {065014} (\bibinfo {year} {2015}{\natexlab{b}})},\ \Eprint
  {http://arxiv.org/abs/1412.0789} {arXiv:1412.0789 [hep-ph]} \BibitemShut
  {NoStop}%
\bibitem [{\citenamefont {Axenides}\ \emph {et~al.}(1983)\citenamefont
  {Axenides}, \citenamefont {Brandenberger},\ and\ \citenamefont
  {Turner}}]{Axenides:1983hj}%
  \BibitemOpen
  \bibfield  {author} {\bibinfo {author} {\bibfnamefont {M.}~\bibnamefont
  {Axenides}}, \bibinfo {author} {\bibfnamefont {R.~H.}\ \bibnamefont
  {Brandenberger}}, \ and\ \bibinfo {author} {\bibfnamefont {M.~S.}\
  \bibnamefont {Turner}},\ }\href {\doibase 10.1016/0370-2693(83)90586-5}
  {\bibfield  {journal} {\bibinfo  {journal} {Phys. Lett. B}\ }\textbf
  {\bibinfo {volume} {126}},\ \bibinfo {pages} {178} (\bibinfo {year}
  {1983})}\BibitemShut {NoStop}%
\bibitem [{\citenamefont {Seckel}\ and\ \citenamefont
  {Turner}(1985)}]{Seckel:1985tj}%
  \BibitemOpen
  \bibfield  {author} {\bibinfo {author} {\bibfnamefont {D.}~\bibnamefont
  {Seckel}}\ and\ \bibinfo {author} {\bibfnamefont {M.~S.}\ \bibnamefont
  {Turner}},\ }\href {\doibase 10.1103/PhysRevD.32.3178} {\bibfield  {journal}
  {\bibinfo  {journal} {Phys. Rev. D}\ }\textbf {\bibinfo {volume} {32}},\
  \bibinfo {pages} {3178} (\bibinfo {year} {1985})}\BibitemShut {NoStop}%
\bibitem [{\citenamefont {Turner}\ and\ \citenamefont
  {Wilczek}(1991)}]{Turner:1990uz}%
  \BibitemOpen
  \bibfield  {author} {\bibinfo {author} {\bibfnamefont {M.~S.}\ \bibnamefont
  {Turner}}\ and\ \bibinfo {author} {\bibfnamefont {F.}~\bibnamefont
  {Wilczek}},\ }\href {\doibase 10.1103/PhysRevLett.66.5} {\bibfield  {journal}
  {\bibinfo  {journal} {Phys. Rev. Lett.}\ }\textbf {\bibinfo {volume} {66}},\
  \bibinfo {pages} {5} (\bibinfo {year} {1991})}\BibitemShut {NoStop}%
\bibitem [{\citenamefont {Ade}\ \emph {et~al.}(2016)\citenamefont {Ade} \emph
  {et~al.}}]{Ade:2015lrj}%
  \BibitemOpen
  \bibfield  {author} {\bibinfo {author} {\bibfnamefont {P.}~\bibnamefont
  {Ade}} \emph {et~al.} (\bibinfo {collaboration} {Planck}),\ }\href {\doibase
  10.1051/0004-6361/201525898} {\bibfield  {journal} {\bibinfo  {journal}
  {Astron. Astrophys.}\ }\textbf {\bibinfo {volume} {594}},\ \bibinfo {pages}
  {A20} (\bibinfo {year} {2016})},\ \Eprint {http://arxiv.org/abs/1502.02114}
  {arXiv:1502.02114 [astro-ph.CO]} \BibitemShut {NoStop}%
\bibitem [{\citenamefont {Kawasaki}\ \emph {et~al.}(2016)\citenamefont
  {Kawasaki}, \citenamefont {Yamada}, \citenamefont {Yanagida},\ and\
  \citenamefont {Yokozaki}}]{Kawasaki:2015mvm}%
  \BibitemOpen
  \bibfield  {author} {\bibinfo {author} {\bibfnamefont {M.}~\bibnamefont
  {Kawasaki}}, \bibinfo {author} {\bibfnamefont {M.}~\bibnamefont {Yamada}},
  \bibinfo {author} {\bibfnamefont {T.~T.}\ \bibnamefont {Yanagida}}, \ and\
  \bibinfo {author} {\bibfnamefont {N.}~\bibnamefont {Yokozaki}},\ }\href
  {\doibase 10.1103/PhysRevD.93.055022} {\bibfield  {journal} {\bibinfo
  {journal} {Phys. Rev. D}\ }\textbf {\bibinfo {volume} {93}},\ \bibinfo
  {pages} {055022} (\bibinfo {year} {2016})},\ \Eprint
  {http://arxiv.org/abs/1512.04259} {arXiv:1512.04259 [hep-ph]} \BibitemShut
  {NoStop}%
\bibitem [{\citenamefont {Masso}\ \emph {et~al.}(2002)\citenamefont {Masso},
  \citenamefont {Rota},\ and\ \citenamefont {Zsembinszki}}]{Masso:2002np}%
  \BibitemOpen
  \bibfield  {author} {\bibinfo {author} {\bibfnamefont {E.}~\bibnamefont
  {Masso}}, \bibinfo {author} {\bibfnamefont {F.}~\bibnamefont {Rota}}, \ and\
  \bibinfo {author} {\bibfnamefont {G.}~\bibnamefont {Zsembinszki}},\ }\href
  {\doibase 10.1103/PhysRevD.66.023004} {\bibfield  {journal} {\bibinfo
  {journal} {Phys. Rev. D}\ }\textbf {\bibinfo {volume} {66}},\ \bibinfo
  {pages} {023004} (\bibinfo {year} {2002})},\ \Eprint
  {http://arxiv.org/abs/hep-ph/0203221} {arXiv:hep-ph/0203221} \BibitemShut
  {NoStop}%
\bibitem [{\citenamefont {Kawasaki}\ \emph
  {et~al.}(2015{\natexlab{c}})\citenamefont {Kawasaki}, \citenamefont
  {Yamada},\ and\ \citenamefont {Yanagida}}]{Kawasaki:2015ofa}%
  \BibitemOpen
  \bibfield  {author} {\bibinfo {author} {\bibfnamefont {M.}~\bibnamefont
  {Kawasaki}}, \bibinfo {author} {\bibfnamefont {M.}~\bibnamefont {Yamada}}, \
  and\ \bibinfo {author} {\bibfnamefont {T.~T.}\ \bibnamefont {Yanagida}},\
  }\href {\doibase 10.1103/PhysRevD.91.125018} {\bibfield  {journal} {\bibinfo
  {journal} {Phys. Rev. D}\ }\textbf {\bibinfo {volume} {91}},\ \bibinfo
  {pages} {125018} (\bibinfo {year} {2015}{\natexlab{c}})},\ \Eprint
  {http://arxiv.org/abs/1504.04126} {arXiv:1504.04126 [hep-ph]} \BibitemShut
  {NoStop}%
\bibitem [{\citenamefont {Turner}(1987)}]{Turner:1986tb}%
  \BibitemOpen
  \bibfield  {author} {\bibinfo {author} {\bibfnamefont {M.~S.}\ \bibnamefont
  {Turner}},\ }\href {\doibase 10.1103/PhysRevLett.59.2489} {\bibfield
  {journal} {\bibinfo  {journal} {Phys. Rev. Lett.}\ }\textbf {\bibinfo
  {volume} {59}},\ \bibinfo {pages} {2489} (\bibinfo {year} {1987})},\ \bibinfo
  {note} {[Erratum: Phys.Rev.Lett. 60, 1101 (1988)]}\BibitemShut {NoStop}%
\bibitem [{\citenamefont {Nakayama}\ \emph {et~al.}(2011)\citenamefont
  {Nakayama}, \citenamefont {Takahashi},\ and\ \citenamefont
  {Yanagida}}]{Nakayama:2010vs}%
  \BibitemOpen
  \bibfield  {author} {\bibinfo {author} {\bibfnamefont {K.}~\bibnamefont
  {Nakayama}}, \bibinfo {author} {\bibfnamefont {F.}~\bibnamefont {Takahashi}},
  \ and\ \bibinfo {author} {\bibfnamefont {T.~T.}\ \bibnamefont {Yanagida}},\
  }\href {\doibase 10.1016/j.physletb.2011.02.013} {\bibfield  {journal}
  {\bibinfo  {journal} {Phys. Lett. B}\ }\textbf {\bibinfo {volume} {697}},\
  \bibinfo {pages} {275} (\bibinfo {year} {2011})},\ \Eprint
  {http://arxiv.org/abs/1010.5693} {arXiv:1010.5693 [hep-ph]} \BibitemShut
  {NoStop}%
\bibitem [{\citenamefont {Weinberg}(2013)}]{Weinberg:2013kea}%
  \BibitemOpen
  \bibfield  {author} {\bibinfo {author} {\bibfnamefont {S.}~\bibnamefont
  {Weinberg}},\ }\href {\doibase 10.1103/PhysRevLett.110.241301} {\bibfield
  {journal} {\bibinfo  {journal} {Phys. Rev. Lett.}\ }\textbf {\bibinfo
  {volume} {110}},\ \bibinfo {pages} {241301} (\bibinfo {year} {2013})},\
  \Eprint {http://arxiv.org/abs/1305.1971} {arXiv:1305.1971 [astro-ph.CO]}
  \BibitemShut {NoStop}%
\bibitem [{\citenamefont {Salvio}\ \emph {et~al.}(2014)\citenamefont {Salvio},
  \citenamefont {Strumia},\ and\ \citenamefont {Xue}}]{Salvio:2013iaa}%
  \BibitemOpen
  \bibfield  {author} {\bibinfo {author} {\bibfnamefont {A.}~\bibnamefont
  {Salvio}}, \bibinfo {author} {\bibfnamefont {A.}~\bibnamefont {Strumia}}, \
  and\ \bibinfo {author} {\bibfnamefont {W.}~\bibnamefont {Xue}},\ }\href
  {\doibase 10.1088/1475-7516/2014/01/011} {\bibfield  {journal} {\bibinfo
  {journal} {JCAP}\ }\textbf {\bibinfo {volume} {01}},\ \bibinfo {pages} {011}
  (\bibinfo {year} {2014})},\ \Eprint {http://arxiv.org/abs/1310.6982}
  {arXiv:1310.6982 [hep-ph]} \BibitemShut {NoStop}%
\bibitem [{\citenamefont {Wu}\ \emph {et~al.}(2014)\citenamefont {Wu},
  \citenamefont {Errard}, \citenamefont {Dvorkin}, \citenamefont {Kuo},
  \citenamefont {Lee}, \citenamefont {McDonald}, \citenamefont {Slosar},\ and\
  \citenamefont {Zahn}}]{Wu:2014hta}%
  \BibitemOpen
  \bibfield  {author} {\bibinfo {author} {\bibfnamefont {W.}~\bibnamefont
  {Wu}}, \bibinfo {author} {\bibfnamefont {J.}~\bibnamefont {Errard}}, \bibinfo
  {author} {\bibfnamefont {C.}~\bibnamefont {Dvorkin}}, \bibinfo {author}
  {\bibfnamefont {C.}~\bibnamefont {Kuo}}, \bibinfo {author} {\bibfnamefont
  {A.}~\bibnamefont {Lee}}, \bibinfo {author} {\bibfnamefont {P.}~\bibnamefont
  {McDonald}}, \bibinfo {author} {\bibfnamefont {A.}~\bibnamefont {Slosar}}, \
  and\ \bibinfo {author} {\bibfnamefont {O.}~\bibnamefont {Zahn}},\ }\href
  {\doibase 10.1088/0004-637X/788/2/138} {\bibfield  {journal} {\bibinfo
  {journal} {Astrophys. J.}\ }\textbf {\bibinfo {volume} {788}},\ \bibinfo
  {pages} {138} (\bibinfo {year} {2014})},\ \Eprint
  {http://arxiv.org/abs/1402.4108} {arXiv:1402.4108 [astro-ph.CO]} \BibitemShut
  {NoStop}%
\bibitem [{\citenamefont {Abazajian}\ \emph {et~al.}(2015)\citenamefont
  {Abazajian} \emph {et~al.}}]{Abazajian:2013oma}%
  \BibitemOpen
  \bibfield  {author} {\bibinfo {author} {\bibfnamefont {K.}~\bibnamefont
  {Abazajian}} \emph {et~al.} (\bibinfo {collaboration} {Topical Conveners:
  K.N. Abazajian, J.E. Carlstrom, A.T. Lee}),\ }\href {\doibase
  10.1016/j.astropartphys.2014.05.014} {\bibfield  {journal} {\bibinfo
  {journal} {Astropart. Phys.}\ }\textbf {\bibinfo {volume} {63}},\ \bibinfo
  {pages} {66} (\bibinfo {year} {2015})},\ \Eprint
  {http://arxiv.org/abs/1309.5383} {arXiv:1309.5383 [astro-ph.CO]} \BibitemShut
  {NoStop}%
\bibitem [{\citenamefont {Armengaud}\ \emph {et~al.}(2014)\citenamefont
  {Armengaud} \emph {et~al.}}]{Armengaud:2014gea}%
  \BibitemOpen
  \bibfield  {author} {\bibinfo {author} {\bibfnamefont {E.}~\bibnamefont
  {Armengaud}} \emph {et~al.},\ }\href {\doibase 10.1088/1748-0221/9/05/T05002}
  {\bibfield  {journal} {\bibinfo  {journal} {JINST}\ }\textbf {\bibinfo
  {volume} {9}},\ \bibinfo {pages} {T05002} (\bibinfo {year} {2014})},\ \Eprint
  {http://arxiv.org/abs/1401.3233} {arXiv:1401.3233 [physics.ins-det]}
  \BibitemShut {NoStop}%
\bibitem [{\citenamefont {Casimir}(1948)}]{Casimir:1948dh}%
  \BibitemOpen
  \bibfield  {author} {\bibinfo {author} {\bibfnamefont {H.~B.~G.}\
  \bibnamefont {Casimir}},\ }\href@noop {} {\bibfield  {journal} {\bibinfo
  {journal} {Indag. Math.}\ }\textbf {\bibinfo {volume} {10}},\ \bibinfo
  {pages} {261} (\bibinfo {year} {1948})}\BibitemShut {NoStop}%
\bibitem [{\citenamefont {Lifshitz}(1956)}]{Lifshitz:1956zz}%
  \BibitemOpen
  \bibfield  {author} {\bibinfo {author} {\bibfnamefont {E.~M.}\ \bibnamefont
  {Lifshitz}},\ }\href@noop {} {\bibfield  {journal} {\bibinfo  {journal} {Sov.
  Phys. JETP}\ }\textbf {\bibinfo {volume} {2}},\ \bibinfo {pages} {73}
  (\bibinfo {year} {1956})}\BibitemShut {NoStop}%
\bibitem [{\citenamefont {Kenneth}\ and\ \citenamefont
  {Klich}(2006)}]{Kenneth:2006vr}%
  \BibitemOpen
  \bibfield  {author} {\bibinfo {author} {\bibfnamefont {O.}~\bibnamefont
  {Kenneth}}\ and\ \bibinfo {author} {\bibfnamefont {I.}~\bibnamefont
  {Klich}},\ }\href {\doibase 10.1103/PhysRevLett.97.160401} {\bibfield
  {journal} {\bibinfo  {journal} {Phys. Rev. Lett.}\ }\textbf {\bibinfo
  {volume} {97}},\ \bibinfo {pages} {160401} (\bibinfo {year} {2006})},\
  \Eprint {http://arxiv.org/abs/quant-ph/0601011} {arXiv:quant-ph/0601011}
  \BibitemShut {NoStop}%
\bibitem [{\citenamefont {Jiang}\ and\ \citenamefont
  {Wilczek}(2019)}]{Jiang:2018ivv}%
  \BibitemOpen
  \bibfield  {author} {\bibinfo {author} {\bibfnamefont {Q.-D.}\ \bibnamefont
  {Jiang}}\ and\ \bibinfo {author} {\bibfnamefont {F.}~\bibnamefont
  {Wilczek}},\ }\href {\doibase 10.1103/PhysRevB.99.125403} {\bibfield
  {journal} {\bibinfo  {journal} {Phys. Rev. B}\ }\textbf {\bibinfo {volume}
  {99}},\ \bibinfo {pages} {125403} (\bibinfo {year} {2019})},\ \Eprint
  {http://arxiv.org/abs/1805.07994} {arXiv:1805.07994 [cond-mat.mes-hall]}
  \BibitemShut {NoStop}%
\end{thebibliography}%

\end{document}